\documentclass[pdftex,epjc3]{svjour3}

\RequirePackage[T1]{fontenc}
\RequirePackage{graphicx}
\RequirePackage{mathptmx}
\RequirePackage{flushend}
\RequirePackage[numbers,sort&compress]{natbib}
\RequirePackage[colorlinks,citecolor=blue,urlcolor=blue,linkcolor=blue]{hyperref}
\usepackage{amsmath,amssymb}
\usepackage{units}

\newcommand{\alphas}{\alpha_\text{s}}
\newcommand{\pt}{p_\perp}

\journalname{Eur. Phys. J. C}

\begin{document}

\title{{\hfill \normalsize \rm{CERN-PH-TH/2013-255, MCnet-13-17}\\ \medskip \\}
JEWEL\,2.0.0 -- Directions for Use}

\author{Korinna Zapp\thanksref{e1,addr1}}

\thankstext{e1}{e-mail: Korinna.Zapp@cern.ch}

\institute{Department of Physics, CERN, Theory Unit, CH-1211 Geneva 23\label{addr1}}

\date{Received: date / Accepted: date}

\maketitle

\begin{abstract}
In this publication the first official release of the \textsc{Jewel\,2.0.0} code\footnote{The first version \textsc{Jewel\,1}~\cite{Zapp:2008gi} could only treat elastic scattering explicitely and the code was never published.}\footnote{The code can be downloaded from the official \textsc{Jewel} homepage \url{jewel.hepforge.org}.} is presented. \textsc{Jewel} is a Monte Carlo event generator simulating QCD jet evolution in heavy-ion collisions. It treats the interplay of QCD radiation and re-scattering in a medium with fully microscopic dynamics in a consistent perturbative framework with minimal assumptions. After a qualitative introduction into the physics of \textsc{Jewel} detailed information about the practical aspects of using the code is given.
\end{abstract}

\section{Introduction}

In heavy-ion collisions at collider energies jets can be reconstructed, as a substantial part of the jet fragments are accessible above the background. This asks for the theoretical description of multi-particle final states that is most easily achieved using Monte Carlo codes. \textsc{Jewel} is a Monte Carlo that describes the QCD evolution of jets in vacuum and in a medium in a perturbative approach. Only the jets are simulated, the underlying event in proton-proton and the remaining (largely soft) event nucleus-nucleus collisions are not included. The physics and performance of the latest version of \textsc{Jewel} have been discussed in detail elsewhere~\cite{Zapp:2012ak}, the aim of this publication is to make the code available and usable. Here, after a qualitative introduction to the physical picture of \textsc{Jewel}, technical aspects relevant for obtaining meaningful results are discussed. As the algorithmic structure of the code is rather complex users are advised not to modify the code. 

\section{Physics of JEWEL}

\textsc{Jewel} simulates jet evolution in a medium invoking a dynamical picture of jet-medium interactions in a consistent perturbative language~\cite{Zapp:2012ak,Zapp:2008gi}. Scattering in the medium is described by $2\to 2$ pQCD matrix elements with parton showers taking into account possible additional radiation. The assumptions underlying the construction of \textsc{Jewel} are that (i) the medium as resolved by the jet consists of a collection of partons, (ii) the dominant effect of soft scattering can be included by an infra-red continuation of the perturbative matrix elements, (iii) the interplay between competing radiative processes is governed by the formation times of the emissions and (iv) the physical picture behind the LPM effect derived in the eikonal limit is valid also in general kinematics. The reasoning behind this approach is to arrive at a description of jet evolution in a medium that is based as far as possible on perturbative QCD and is minimal in its assumptions.

\textsc{Jewel} implements a fully microscopic description of jet-evolution in a medium including coherence effects. This comes at the price of complexity. The aim of this section is to give a qualitative introduction focussed on the main ideas and a flavour of the performance of the code. For a more formal and detailed discussion of the physics, the implementation and results including uncertainties the reader is referred to the original publication~\cite{Zapp:2012ak}.

\subsection{Qualitative discussion of the physics}

\begin{figure}[ht]
\centering
\includegraphics[scale=0.4]{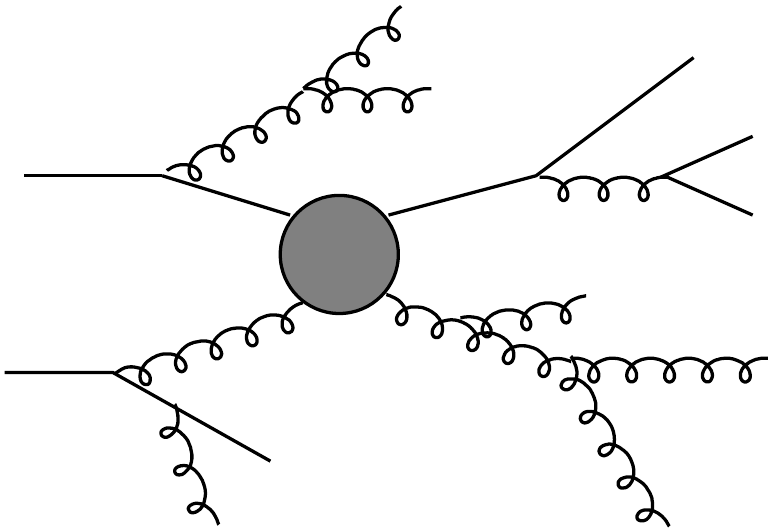}
\caption{Schematic picture of extra emissions generated by the parton shower on top of a hard quark-gluon scattering event described by a $2\to 2$ matrix element and depicted by the shaded blob.}
\label{fig::vac-ps}
\end{figure}

In the absence of the medium \textsc{Jewel} reduces to an ordinary virtuality ordered parton shower similar to the virtuality ordered shower in \textsc{Pythia}\,6~\cite{Sjostrand:2006za}. In QCD hard scattering processes are described by matrix elements at fixed order in perturbation theory. For the discussion here it is sufficient to consider only the lowest order scattering processes, which are the tree level $2\to 2$ processes\footnote{It is well known how to include higher order corrections in fixed order calculations and Monte Carlo event generators, but this is currently not relevant for the discussion of jet quenching and will therefore not be discussed here.}. However, radiative corrections can be large and often need to be taken into account. The leading contribution of radiative corrections has a simple structure and is universal, i.e.\ it does not depend on the kind of hard scattering under consideration. This allows one to systematically construct approximations to the full higher order, i.e.\ $2\to 3$, $2\to 4$ etc., matrix elements. In Monte Carlo event generators this is achieved by first generating a hard scattering configuration from the $2\to 2$ matrix elements and then adding the leading radiative corrections with a parton shower, which attaches extra emissions to all incoming and outgoing legs of the hard scattering. This is sketched in figure~\ref{fig::vac-ps} for the example of a hard quark-gluon scattering event depicted by the shaded blob.

The parton shower is unitary, i.e.\ it does not affect the cross section, and it generates any number of extra emissions or, phrased in a more technical language, it resums the leading (and some of the sub-leading) logarithmic contributions to all orders. The emissions are ordered in a variable that characterises their hardness (for instance the transverse momentum of the emission, or the virtuality), in the initial state the hardness increases until the scale of the matrix element is reached, in the final state it decreases. In the infra-red region the probability for gluon emission diverges and the parton shower thus has to be cut off at a suitable scale. This makes physical sense, as very soft or very collinear emissions will always end up in the same hadron as the emitting parton and are therefore not observable.

In evolving from an infra-red scale to the scale of the hard process in the initial state the parton shower does nothing but an explicit DGLAP evolution. As the emitted partons form the proton structure at different scales the action of the parton shower in the initial state is constrained by the proton PDFs.

To sum up, a parton shower is a theoretically well controlled tool that generates the leading radiative corrections to any scattering process to all orders. In doing so it systematically approximates the higher order $2\to n$ matrix elements. 

\medskip

The matrix element and final state parton shower don't have any knowledge about the origin of the partons they are dealing with. The initial state parton shower only knows through the PDF that the partons originate from a hadron of a certain structure. The only difference between hard partonic scattering in a proton-proton collision and the hard re-scattering of a hard parton off a constituent of a strongly interacting medium is that in the latter case the incoming partons are not part of a proton. Following standard factorisation approaches one can argue that a hard momentum transfer will resolve the partonic structure of any QCD medium irrespective of its behaviour at low scales. It is, however, not a priori clear that the condition of being sufficiently hard is fulfilled for the average interaction of a jet in the medium created in ultra-relativistic heavy-ion collisions. It is assumed to be the case in \textsc{Jewel} (assumption (i)), but only comparison to data will tell to what extent it is justified. 

Although there is no proven factorisation theorem for this case, it seems plausible that for perturbatively hard momentum transfers one can use exactly the same technology of matrix elements and parton showers used for proton-proton collisions to describe the re-scattering of a parton in a medium. Only the PDFs have to be substituted by appropriate 'partonic PDFs' encoding the information about the QCD evolution of partons that are not constituents of a hadron.

In this way radiative corrections to re-scattering in the medium giving rise to radiative energy loss are automatically included (to leading logarithmic accuracy) to all orders. And -- what is equally important -- they are generated with the (leading log) correct relative rates, which is not the case when one naively adds $2\to 3$ matrix elements by hand.

Using matrix elements and parton showers to describe the perturbatively hard re-scattering of a parton in a medium thus appears to be very reasonable. One complication arising in this case is that there is no natural infra-red cut-off. In proton-proton collisions the matrix element is usually guaranteed to be sufficiently hard due to the requirements of the analysis. In the case of jet production, for instance, the jets will be required to have a certain $\pt$. In the case of re-scattering in the medium this is not the case. Very soft momentum transfers will obvioulsy not lead to any visible effect, but it is unclear how the regime between these extremely soft and perturbatively hard momentum transfers should be treated. Here, assumption (ii) of \textsc{Jewel} comes into play: It is assumed that this can be achieved by a suitable infra-red continuation of the matrix elements. For the parton shower nothing changes, as the requirement that the shower should only emit resolvable radiation is still sensible in the context of re-scattering in a medium.  

\textsc{Jewel} thus uses the same language and techniques to describe the initial production of jets and their rescattering in a medium. This allows for a consistent treatment of the entire jet evolution, as will be discussed in the rest of this section.

\medskip

\begin{figure}[ht]
\centering
\includegraphics[width=0.5\linewidth]{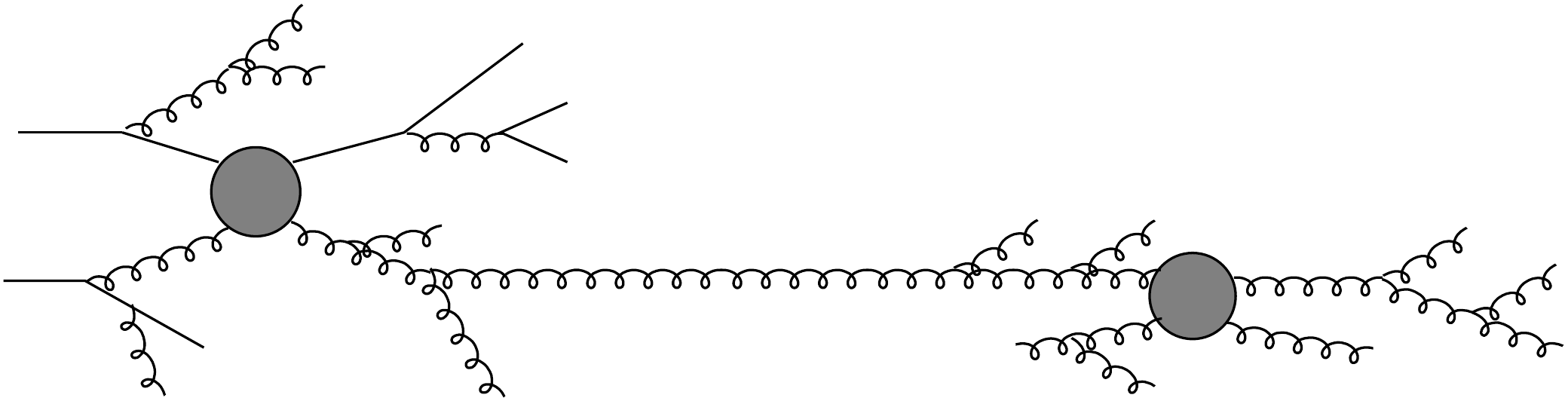}
\caption{Schematic picture of extra emissions in two well separated scattering events. Again, matrix elements are depicted by the shaded blob. The re-scattering is only indicated for one parton, but of course all produced partons can undergo re-scattering in the same way.}
\label{fig::med-ps-1}
\end{figure}

\begin{figure}[ht]
\centering
\includegraphics[width=0.5\linewidth]{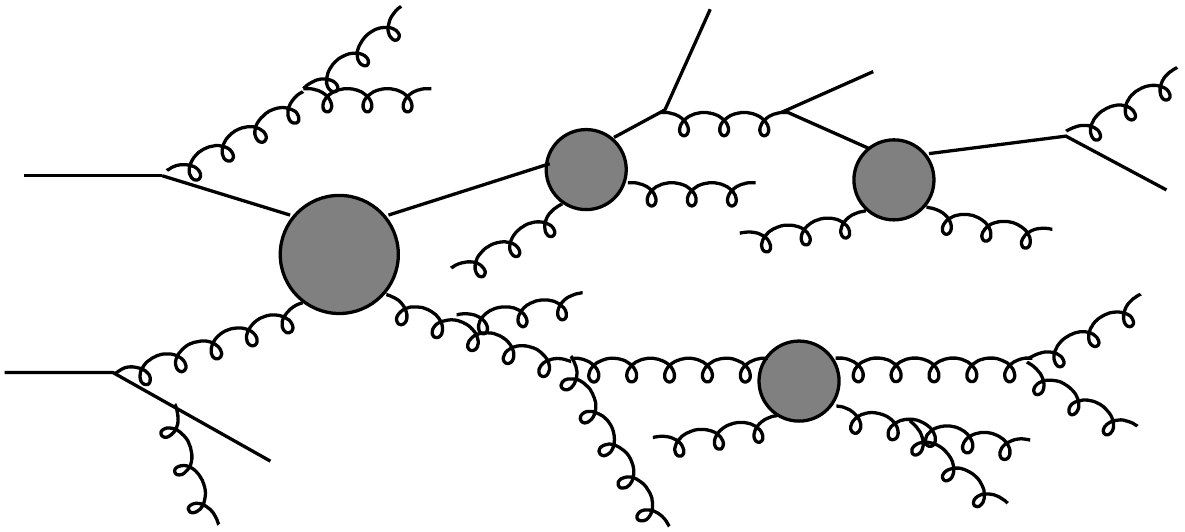}
\caption{Schematic picture of extra emissions and re-scatters taking place on comparable time scales.}
\label{fig::med-ps-2}
\end{figure}

So far it was assumed that the parton re-scattering in a medium is on-shell, which means that the distance between the inial jet production and the first re-scattering as well as between subsequent re-scatters is large compared to the time needed for the parton shower evolution. This situation is sketched in figure~\ref{fig::med-ps-1}. Given that the initial jet production happens in the same nuclear collision as the formation of the medium and that radiation during parton shower evolution does not happen instantaneously but with a certain formation time, this is not necessarily the case. In reality there may be several emissions happening at the same time because the parton shower of the intial hard scattering producing the event has not terminated by the time of the re-scattering and/or two re-scatterings happen at a distance that is shorter than the time needed for the parton shower evolution. As all emissions are handled in exactly the same way in \textsc{Jewel} a formation time can be assigned to all of them consistently. In case of two emissions taking place at the same time the emission with the shorter formation time gets formed while the other one is discarded. In practice, an emission is only formed as an individual parton at the end of the formation time. Before that it is treated as a potential emission that whose formation time is compared to other potential emission. The one with the shortest formation gets formed as a parton while the others are discarded. This situation is sketched in figure~\ref{fig::med-ps-2}. Again, this is to some degree an assumption (cf. assumption (iii)) as it is very difficult to show from first principles that this is the correct treatment, plausible as it may seem. This procedure ensures that re-scatters that are hard compared to the virtuality of the incoming parton will reset the parton shower to starting conditions determined by the kinematics of the re-scattering while soft re-scatterings are unable to induce extra radiation.

\medskip

\begin{figure}[ht]
\centering
\includegraphics[width=0.5\linewidth]{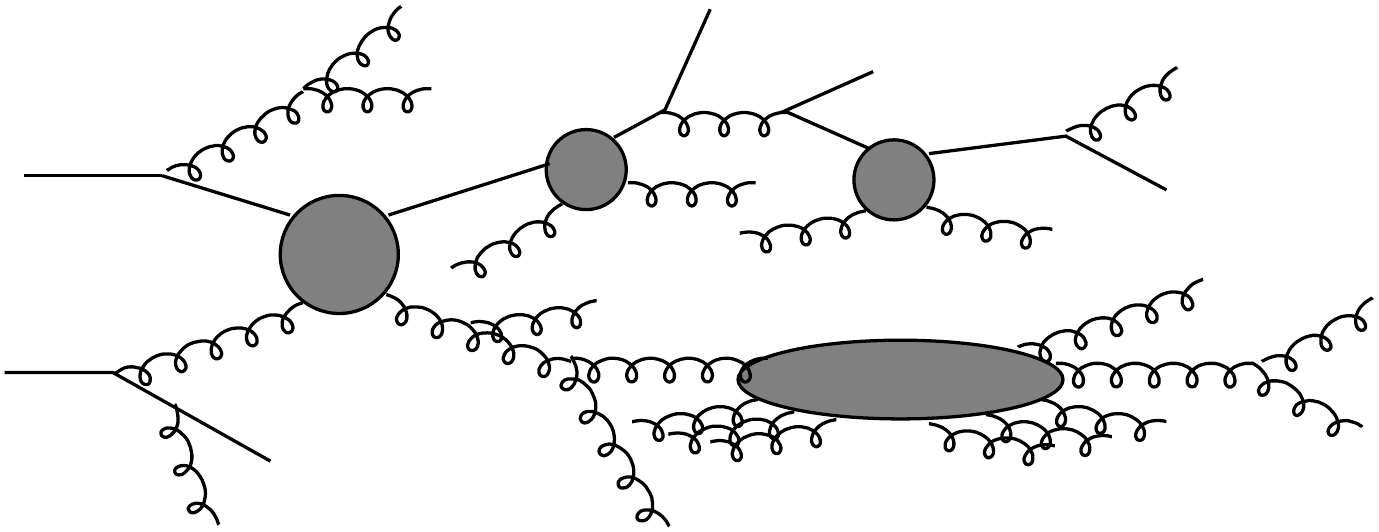}
\caption{Schematic picture of extra radiation and re-scattering where several momentum transfers can act coherently to induce an emission.}
\label{fig::med-ps-3}
\end{figure}

It is known from analytical calculations of bremsstrahlung induced by multiple scattering that radiation induced by subsequent scatterings interferes destructively when the formation times overlap (LPM-effect). Scattering centres within the formation time of the emitted gluon act coherently, only the sum of the individual momentum transfers is relevant for the gluon emission. In addition to changing the distribution of radiated gluons this also has an effect on the emission rate. The LPM-effect can be dealt with in a probabilistic formulation~\cite{Zapp:2008af,Zapp:2011ya} by using an iterative algorithm to determine the formation time of the emission and the coherently contributing momentum tranfers. Together with proper reweighting of the emission this procedure reproduces the analytical results. No analytic results are known for general, i.e. non-eikonal, kinematics and situations with competing sources of radiation. In \textsc{Jewel} it is assumed that the physical picture derived in the eikonal limit is still valid so that the probabilistic algorithm, adapted for the different kinematics, can still be used (assumption (iv)). This essentially means that in some cases a momentum transfer is replaced by the effective momentum transfer from several coherent scatterings (as sketched in figure~\ref{fig::med-ps-3}) and emissions have to be rejected with a certain probability (for details see~\cite{Zapp:2011ya,Zapp:2012ak}).

\medskip

\textsc{Jewel} makes no assumptions about the nature of the medium, but needs to be provided with the phase space density of scattering centres. The results of~\cite{Zapp:2012ak} and section~\ref{sec::results} were obtained with a simple hydrodynamical model~\cite{Zapp:2005kt,Zapp:2012ak}, which describes the boost-invariant longitudinal expansion~\cite{Bjorken:1982qr} of an ideal quark-gluon gas. The transverse profile is fixed by assuming that the energy density is proportional to the density of wounded nucleons in the transverse plane, which is calculated in a Glauber model~\cite{Eskola:1988yh}. Given the initial condition (cf. section~\ref{sec::medmodel}) the density of scattering centres at any space-time point needed for the evaluation of the local scattering rate is easily calculated. When a scattering takes place a scattering centre is generated with a momentum given by the thermal distribution at the local temperature.

\subsection{Some results obtained with JEWEL}
\label{sec::results}

In this section some results will be presented\footnote{All the analyses and plots were done with Rivet~\cite{Buckley:2010ar}.}. The validation and discussion of measurements in $e^++e^-$ and $p+p$ will not be repeated, as they were already discussed in~\cite{Zapp:2012ak}. For the jet evolution in a medium the most important changes are improvements of the medium model (the longitudinal expansion is now properly taken into account also in the momentum distribution of the scattering centres) and a more consistent implementation of the LPM effect, which leads to a reduction of the temperature required to reproduce the experimentally observed jet suppression. The results shown here where obtained with a slightly reduced infra-red regulator $\mu_\text{D} = 0.9\cdot 3T$ and an average initial temperature of $T_\text{i} = \unit[360]{MeV}$ (which corresponds to a central temperature of $T_\text{i} = \unit[486]{MeV}$ for $b = 0$) at $\tau_\text{i} = \unit[0.6]{fm}$. With these values more sophisticated hydrodynamic calculations reproduce the data on soft particle production~\cite{Shen:2012vn}. 

\begin{figure*}[ht]
\centering
\includegraphics[width=0.45\linewidth]{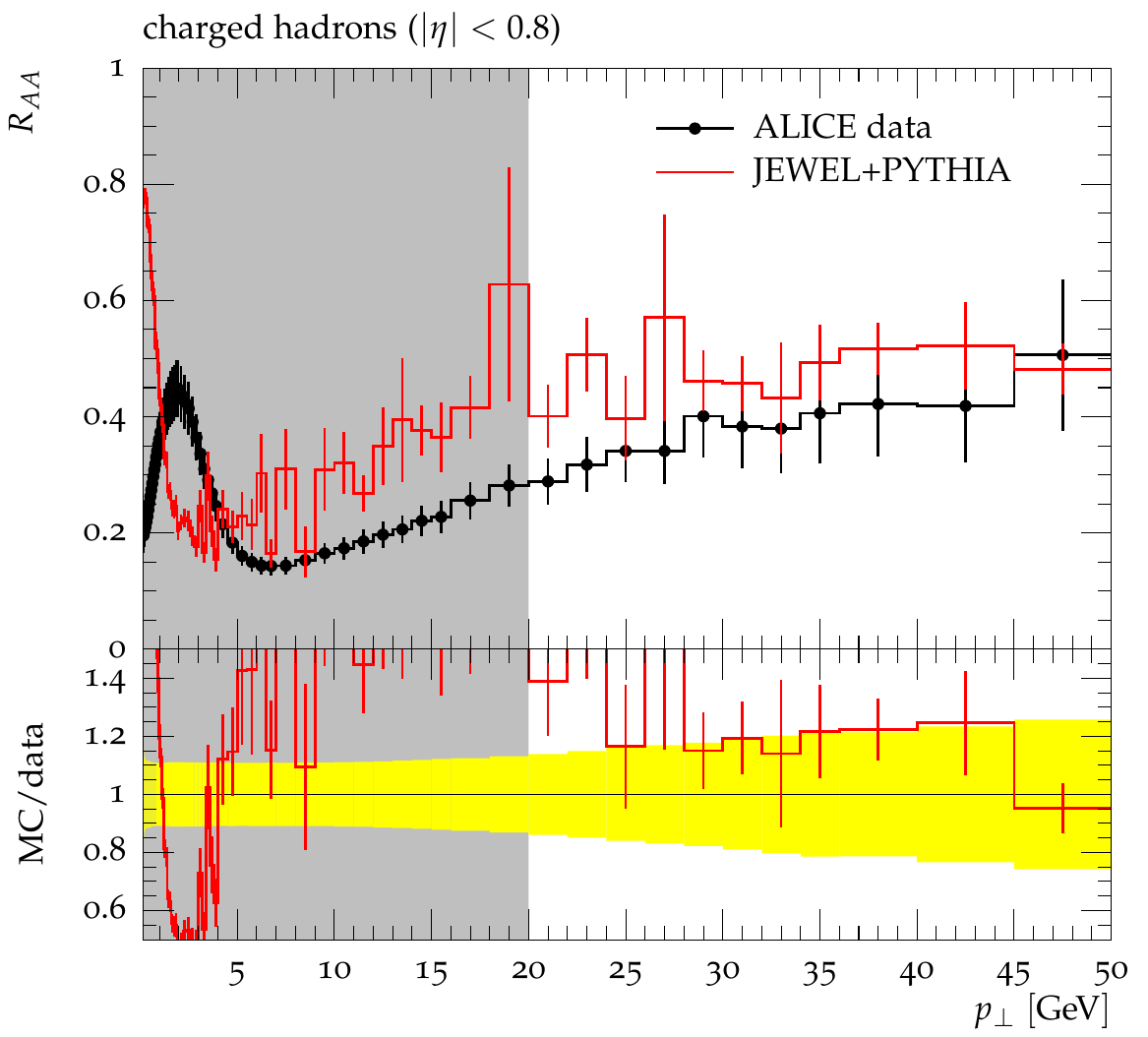}
\includegraphics[width=0.45\linewidth]{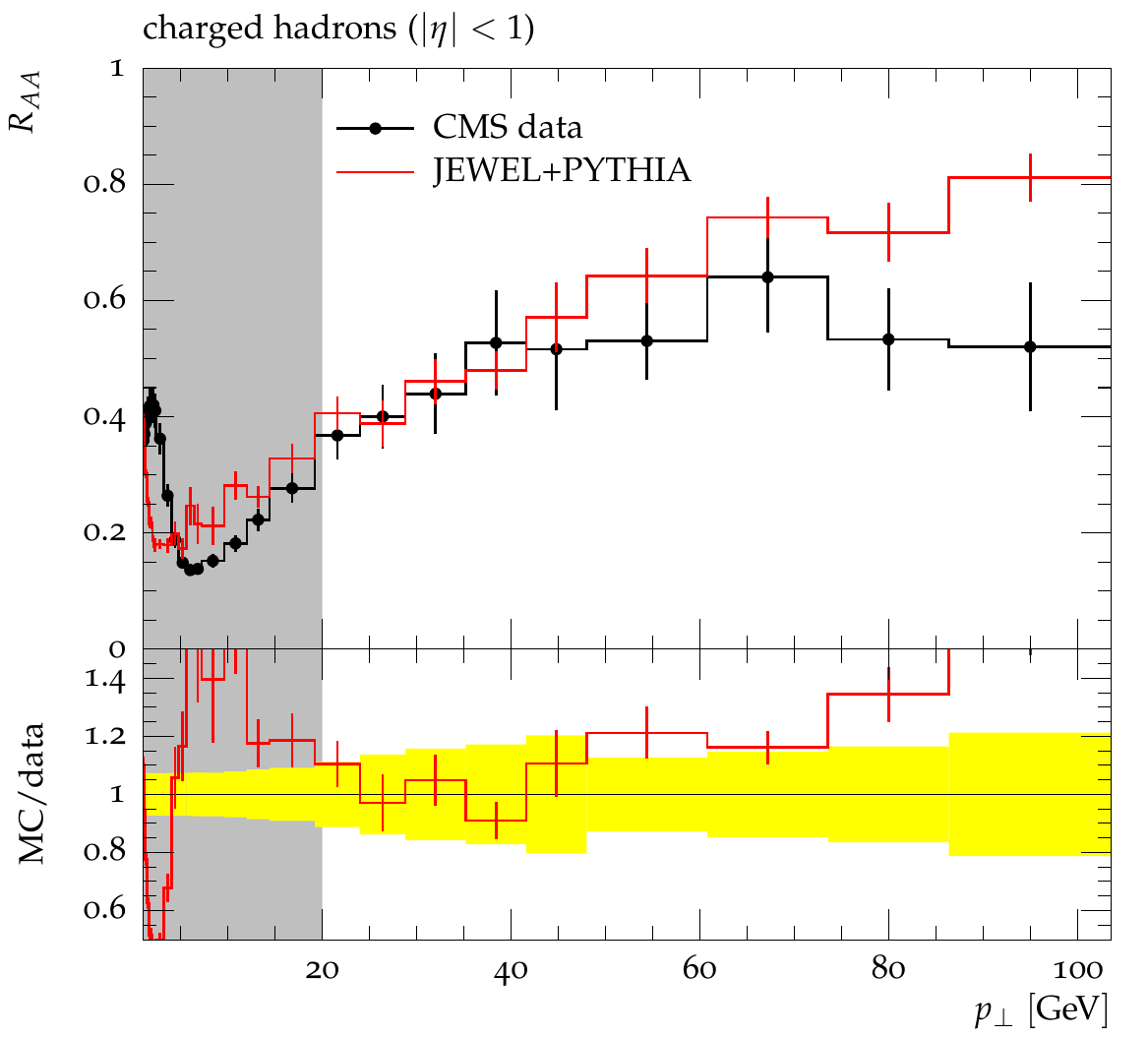}
\caption{Nuclear modification factor for charged hadrons in Pb+Pb collisions at
$\sqrt{s}_\text{NN}=\unit[2.76]{TeV}$ in the \unit[0-5]{\%} and \unit[0-10]{\%} centrality class simulated with \textsc{Jewel+Pythia} and compared to CMS~\cite{CMS:2012aa} and ALICE~\cite{:2012eq} data.}
\label{fig::hadronraa}
\end{figure*}

For $\pt \gtrsim \unit[20]{GeV}$, where the \textsc{Jewel+Pythia} results can be trusted, the nuclear modification factor for hadrons (figure~\ref{fig::hadronraa}) is in reasonable agreement with the data from ALICE and CMS. 

Jets are reconstructed using the same jet algorithm as the experiments, namely the anti-$k_\perp$ algorithm provided by the FastJet package~\cite{Cacciari:2011ma}. However, the comparison of jet observables to data suffers from a slight mismatch between the background subtraction procedures. In data, background is subtracted from the reconstructed jets. Since \textsc{Jewel} does not simulate the soft event it is not possible to follow the same prescription in analysing the Monte Carlo events. Instead, the recoiling scattering centres are removed from the event before hadronisation and no background subtraction is performed. The related uncertainties cannot be estimated without a Monte Carlo model for the entire medium.

\begin{figure*}[ht]
\centering
\includegraphics[width=0.45\linewidth]{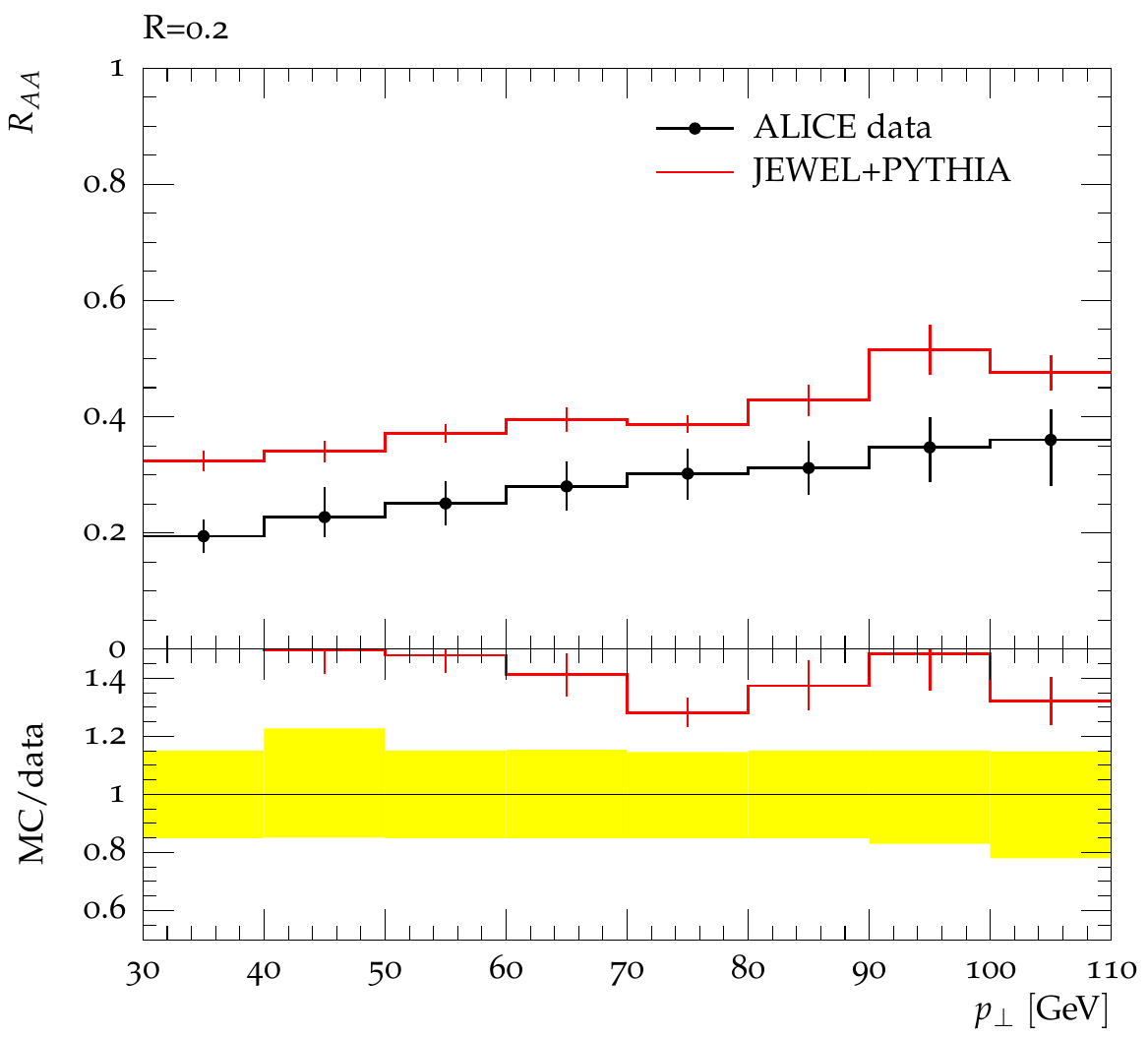}
\includegraphics[width=0.45\linewidth]{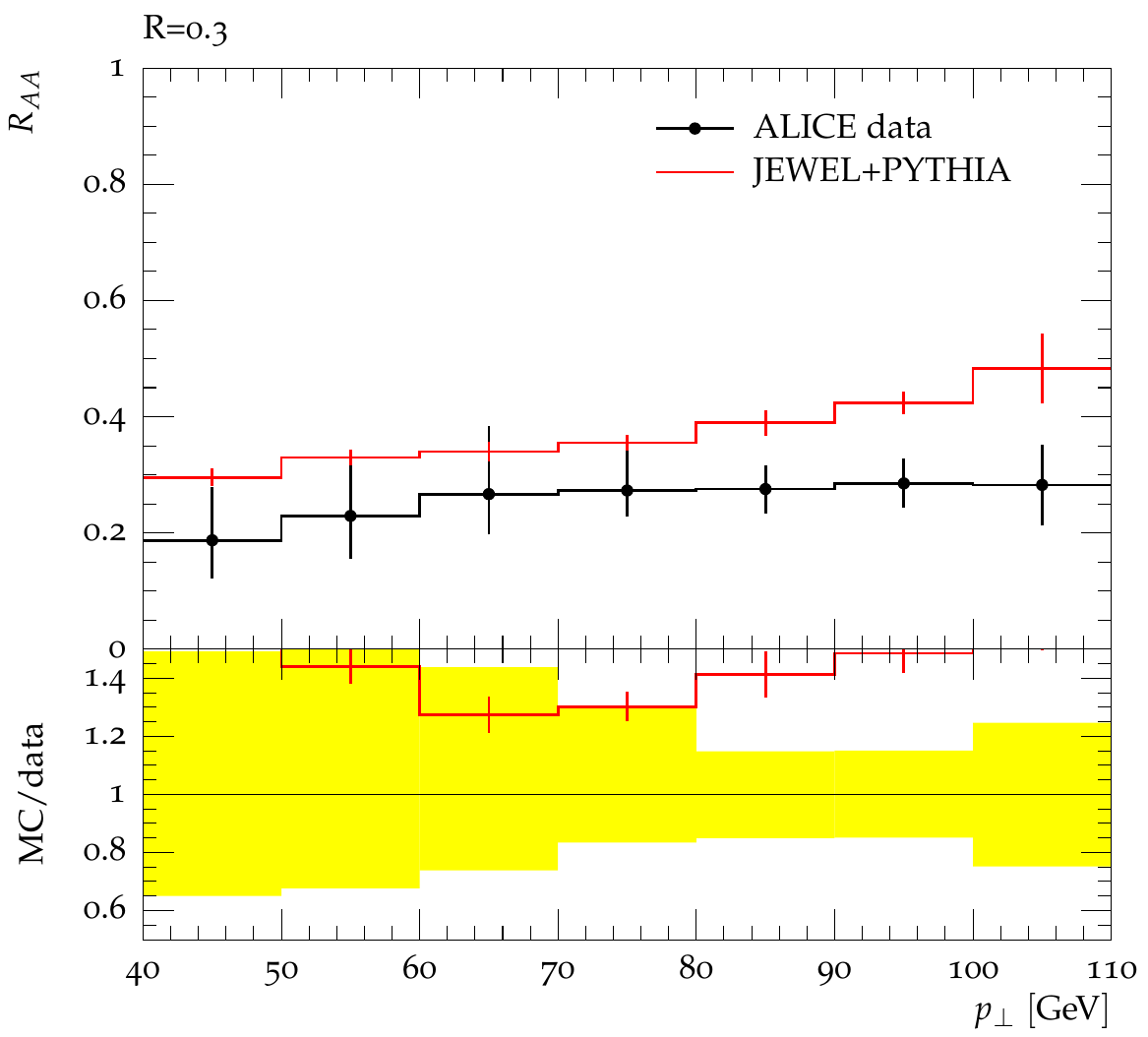}
\caption{\textsc{Jewel+Pythia} results for $R_\text{AA}$ of jets in Pb+Pb collisions at $\sqrt{s}_\text{NN}=\unit[2.76]{TeV}$ compared to ALICE data~\cite{:2012ch} for two values of the jet radius (correlated systematic errors not shown).}
\label{fig::jetraa}
\end{figure*}

\begin{figure*}[ht]
\centering
\includegraphics[width=0.45\linewidth]{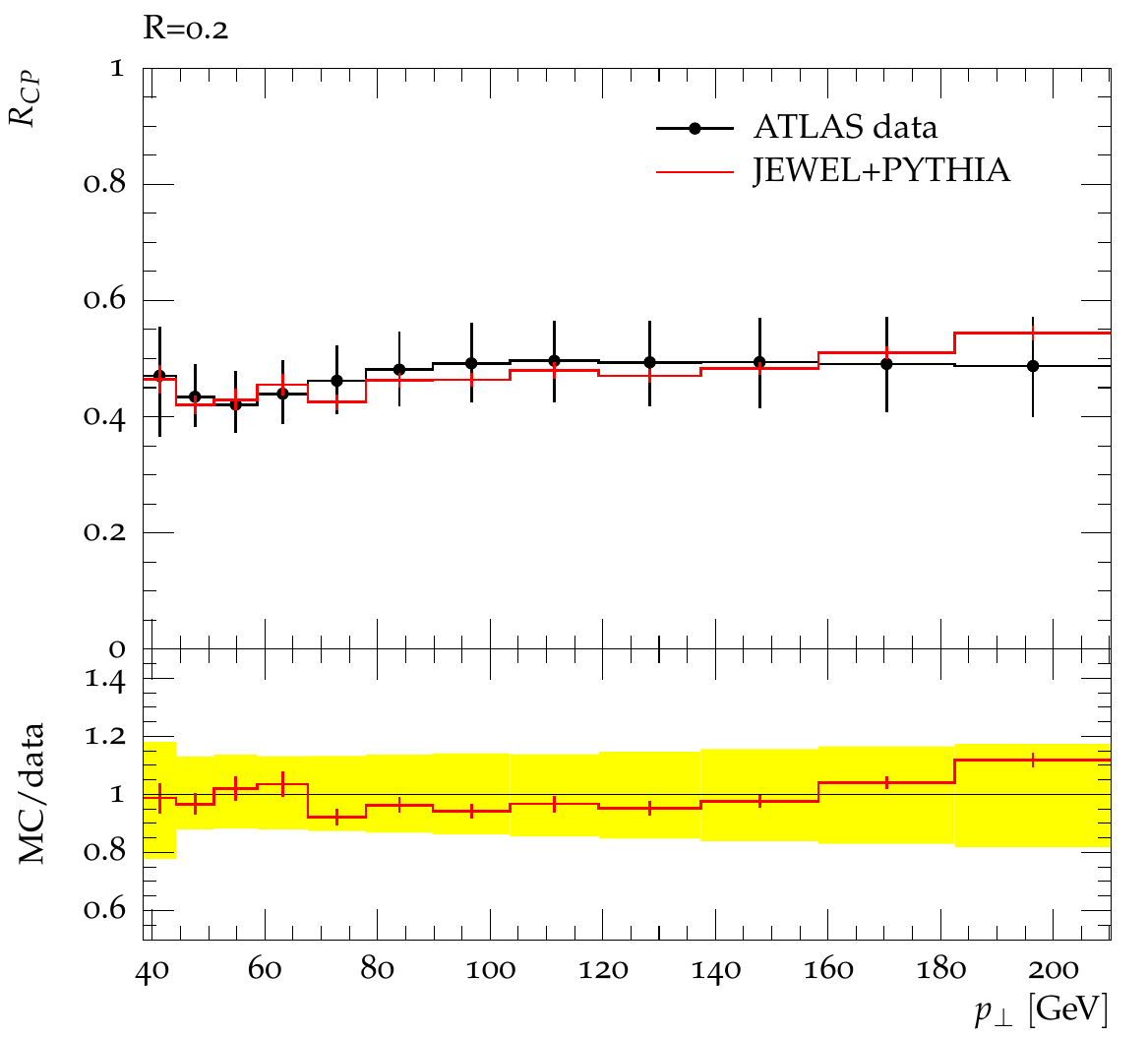}
\includegraphics[width=0.45\linewidth]{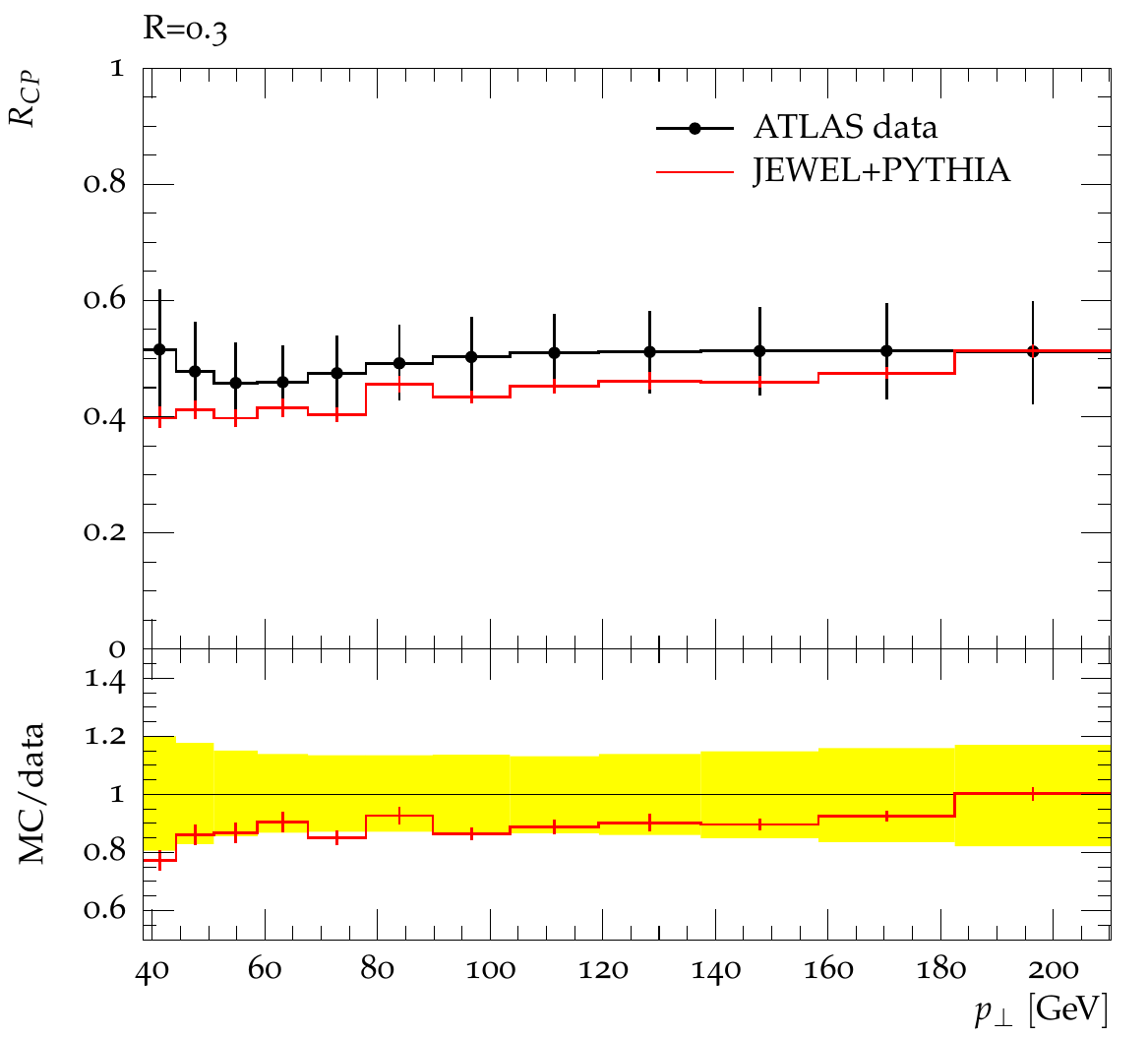}\par
\includegraphics[width=0.45\linewidth]{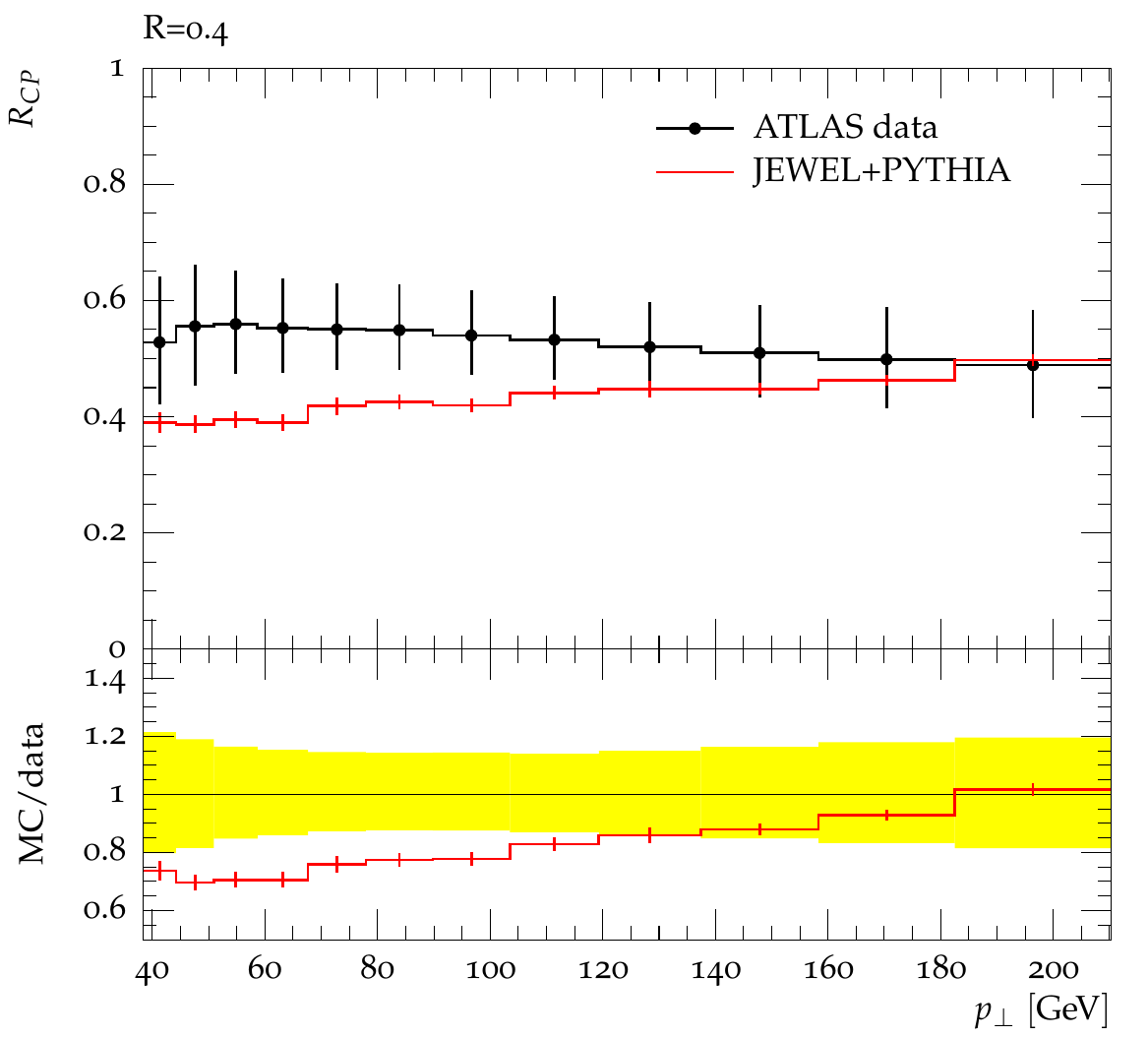}
\includegraphics[width=0.45\linewidth]{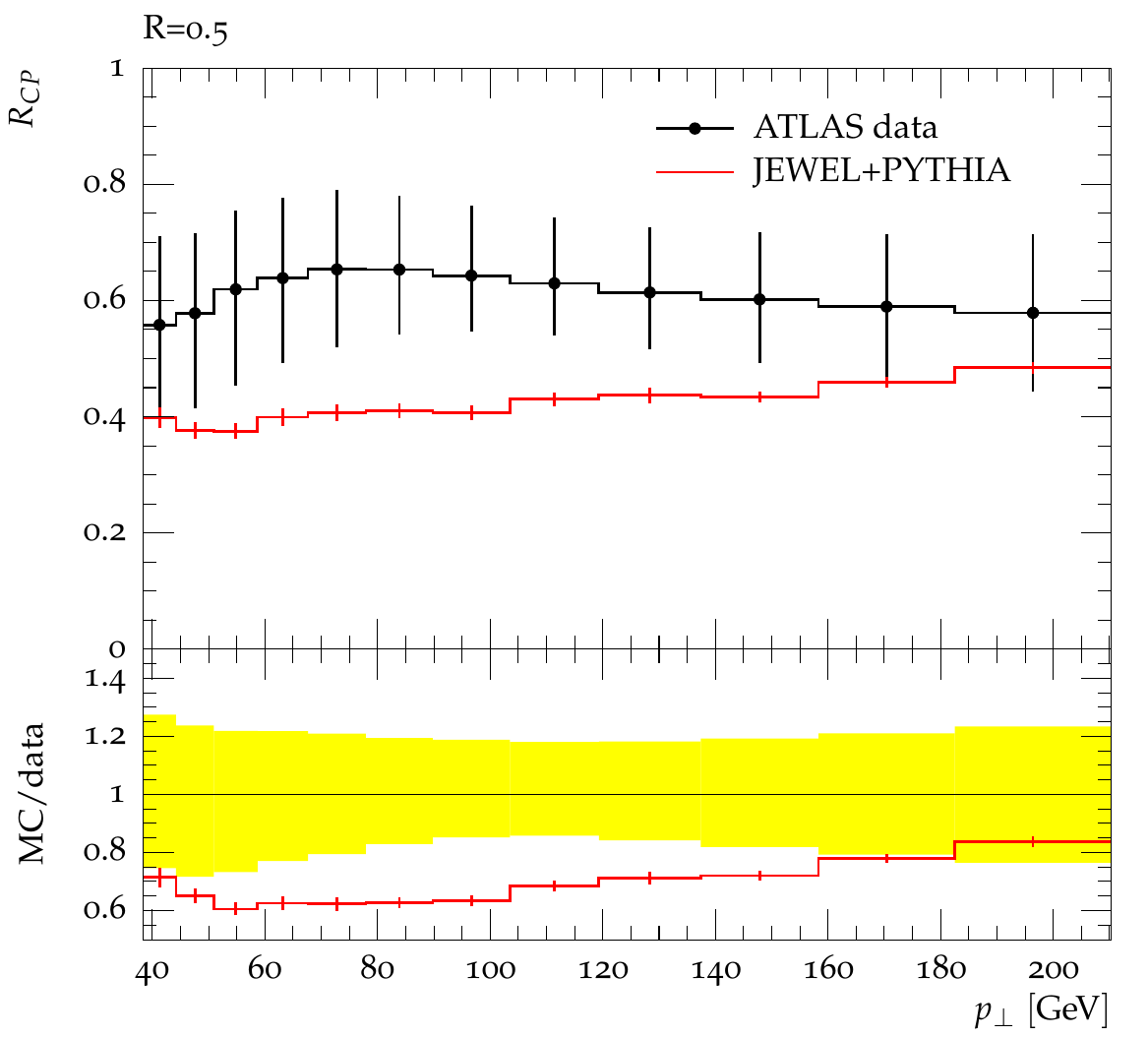}
\caption{\textsc{Jewel+Pythia} results for $R_\text{CP}$ of jets in Pb+Pb collisions at $\sqrt{s}_\text{NN}=\unit[2.76]{TeV}$ compared to ATLAS data~\cite{Aad:2012vca} for different values of the jet radius. The ratio is taken between the \unit[0-10]{\%} and the \unit[60-80]{\%} centrality class.}
\label{fig::jetrcp}
\end{figure*}

The nuclear modification factor for jets (figure~\ref{fig::jetraa}) are slightly larger in the Monte Carlo than in the ALICE data. The values for $R_\text{CP}$ (figure~\ref{fig::jetrcp}), on the other hand, agree well with the ATLAS data for small jet radii, but deviate from the data for larger radii. This suggests that the discrepancy may be due to the different treatment of the background. 

\begin{figure*}[ht]
\centering
\includegraphics[width=0.45\linewidth]{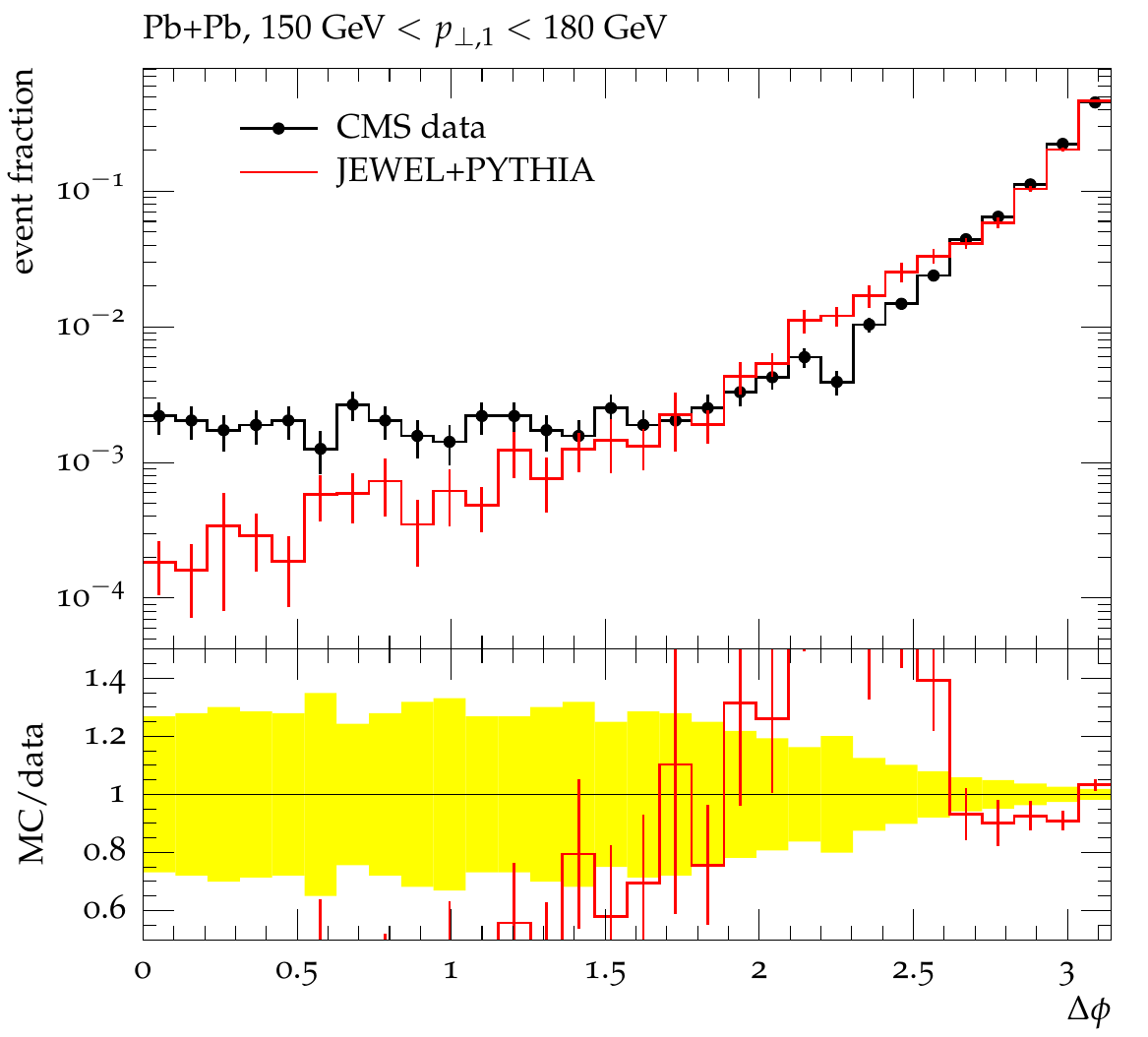}
\includegraphics[width=0.45\linewidth]{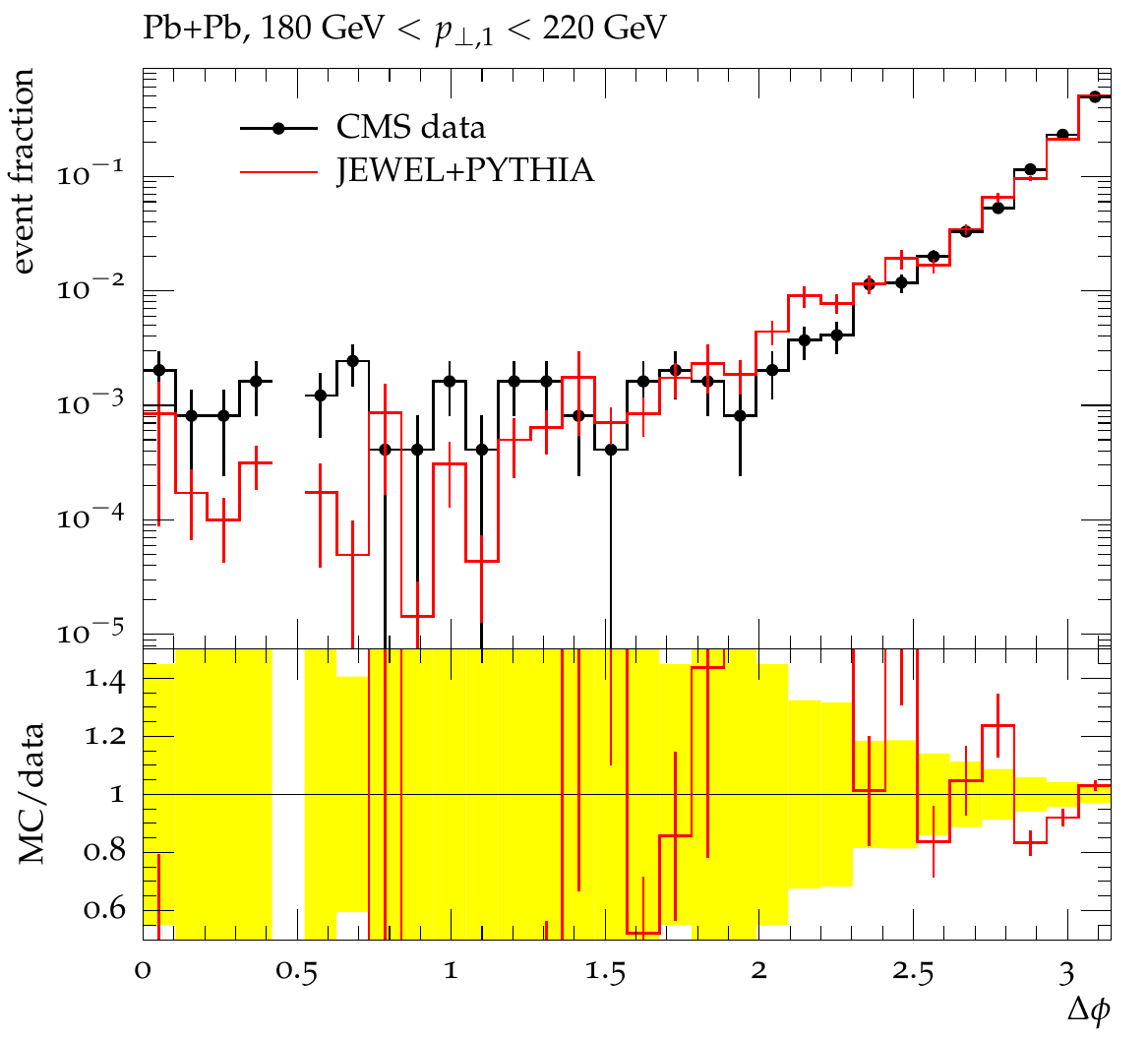}\par
\includegraphics[width=0.45\linewidth]{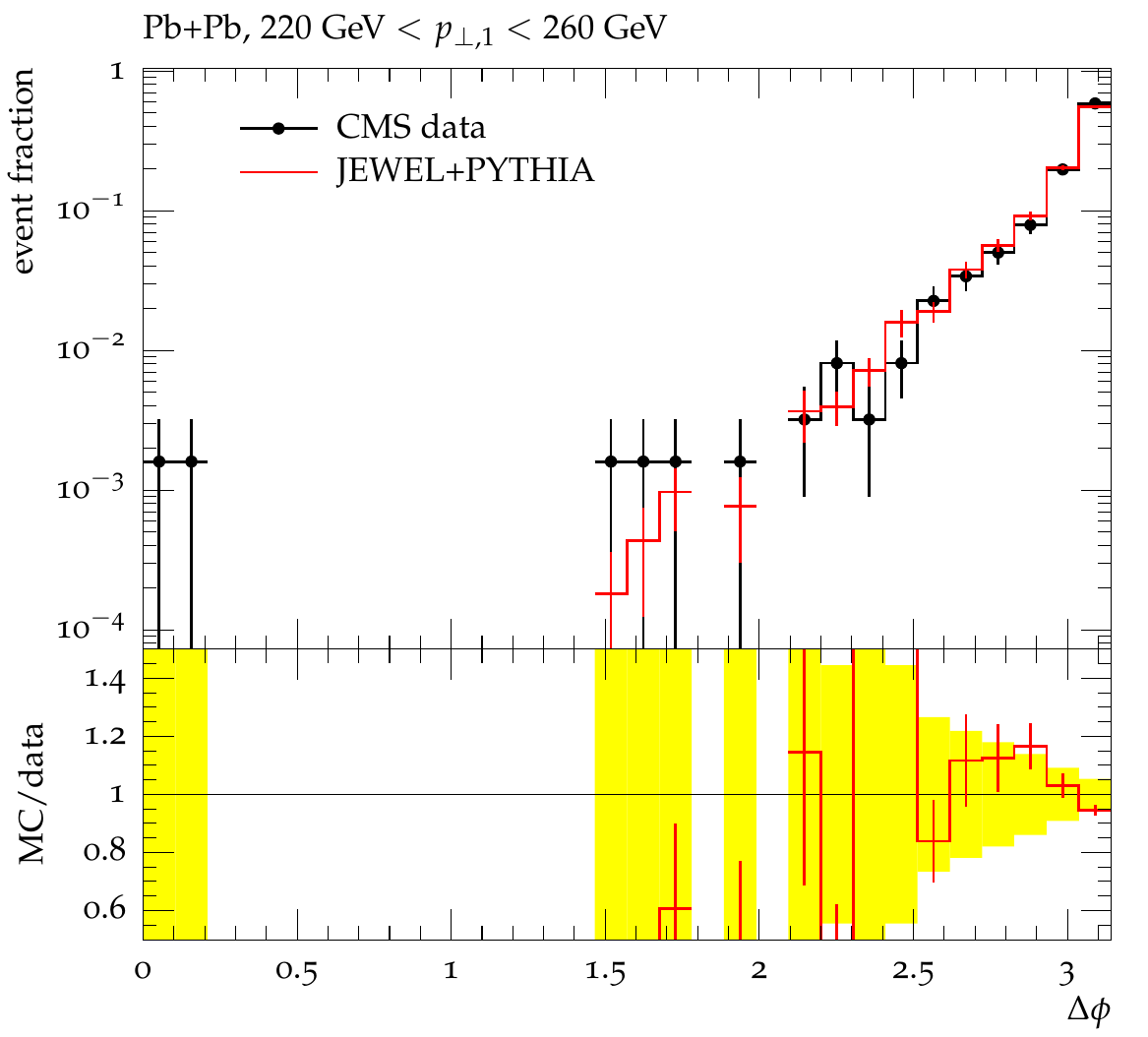}
\includegraphics[width=0.45\linewidth]{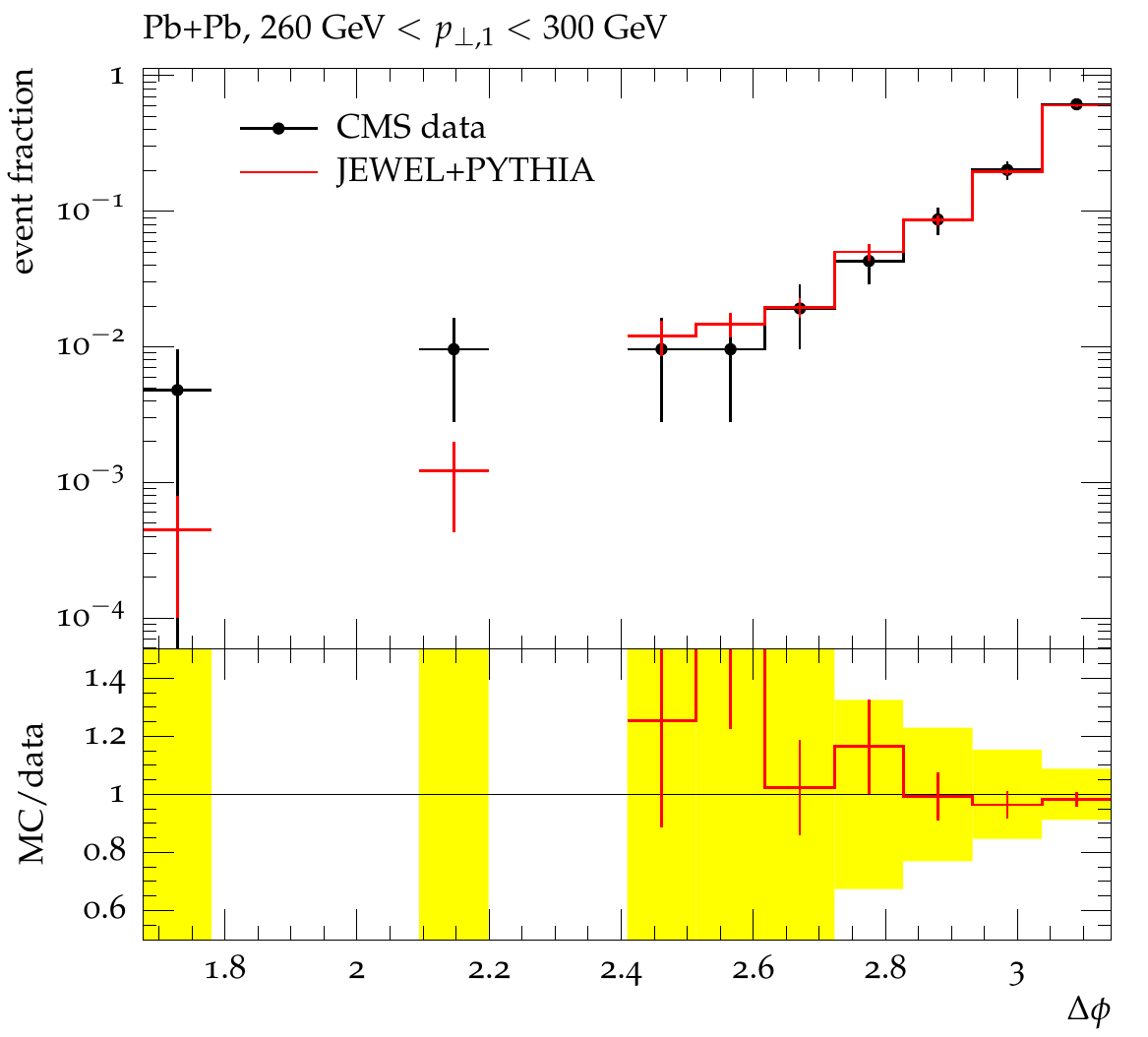}
\caption{\textsc{Jewel+Pythia} results for the azimuthal separation $\Delta \phi$ between two jets in dijet events compared to CMS results~\cite{Chatrchyan:2012nia} in the \unit[0-20]{\%} centrality class. Results are shown for different transverse momenta of the leading jet, the sub-leading jet is required to have $p_{\perp,2} > \unit[30]{GeV}$. The data are not unfolded for jet energy resolution, so the Monte Carlo events were smeared with the parametrisation from~\cite{Chatrchyan:2012gt}.}
\label{fig::deltaphi}
\end{figure*}

CMS has performed a number of measurements characterising di-jet events in Pb+Pb collisions. The comparison of Monte Carlo results to these data is complicated by the fact that the data are not unfolded for the jet energy resolution. The $\pt$ of the jets reconstructed from Monte Carlo events are smeared with the parametrisation of the experimental resolution determined in $\gamma$-jet events~\cite{Chatrchyan:2012gt}. Due to the dominance of quark jets and the possibly different kinematics in this sample the resolution may be different in di-jet events, but CMS did not publish the resolution in pure QCD events.

Figure~\ref{fig::deltaphi} shows the azimuthal decorrelation of di-jets in central Pb+Pb collisions in bins of the transverse momentum $p_{\perp,1}$ of the leading jet (the subleading jet has $p_{\perp,2} > \unit[30]{GeV}$). The \textsc{Jewel+Pythia} results are lacking some support in the region of small $\Delta \phi$, but are otherwise in reasonable agreement with the CMS data. At small $\Delta \phi$ contaminations from fake jets (if there are any) are most visible, which are not present in the Monte Carlo sample. This may be an explanation for the discrepancy, but again this cannot be verified without full modelling of the background.

\begin{figure*}[ht]
\centering
\includegraphics[width=0.45\linewidth]{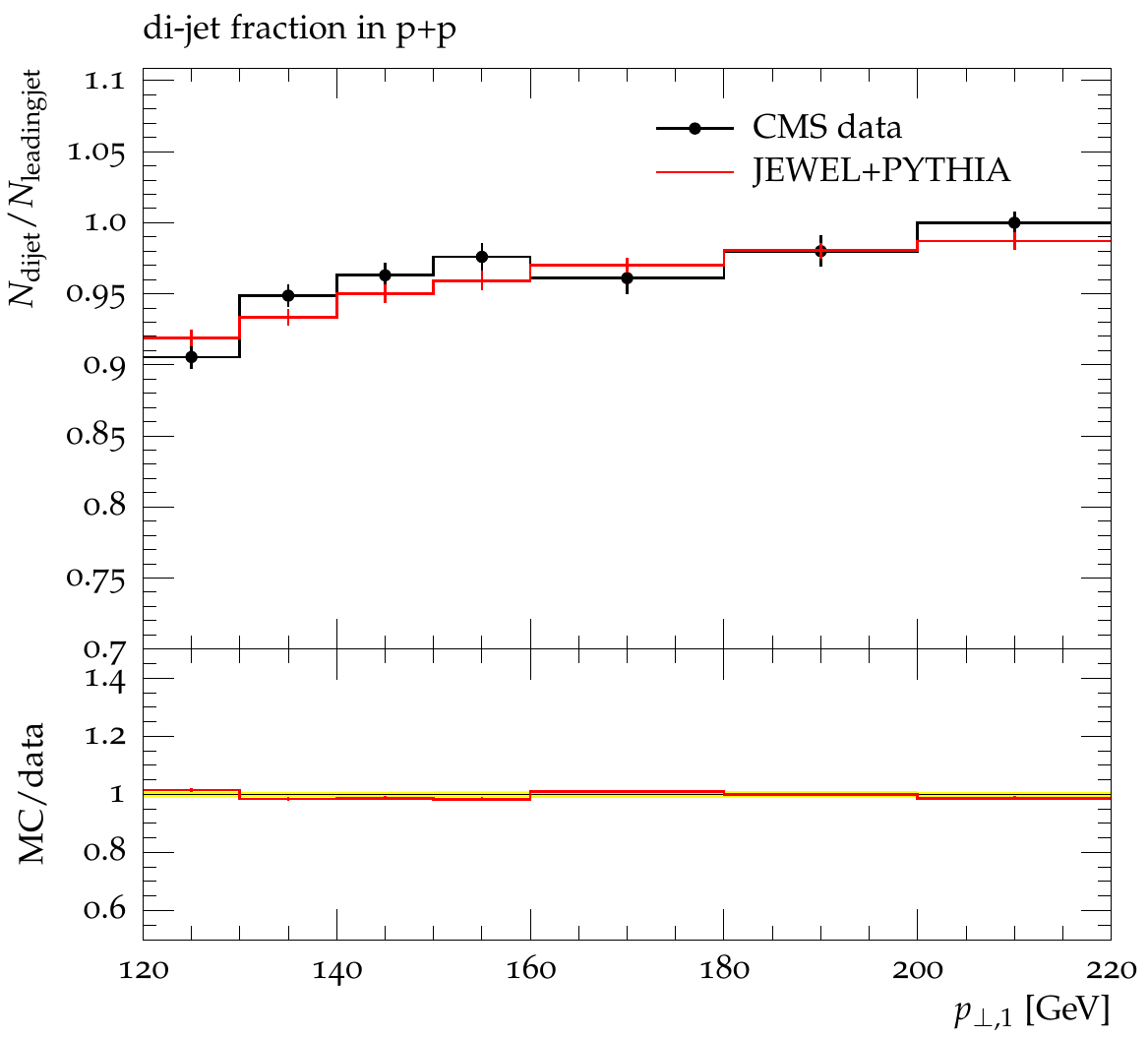}
\includegraphics[width=0.45\linewidth]{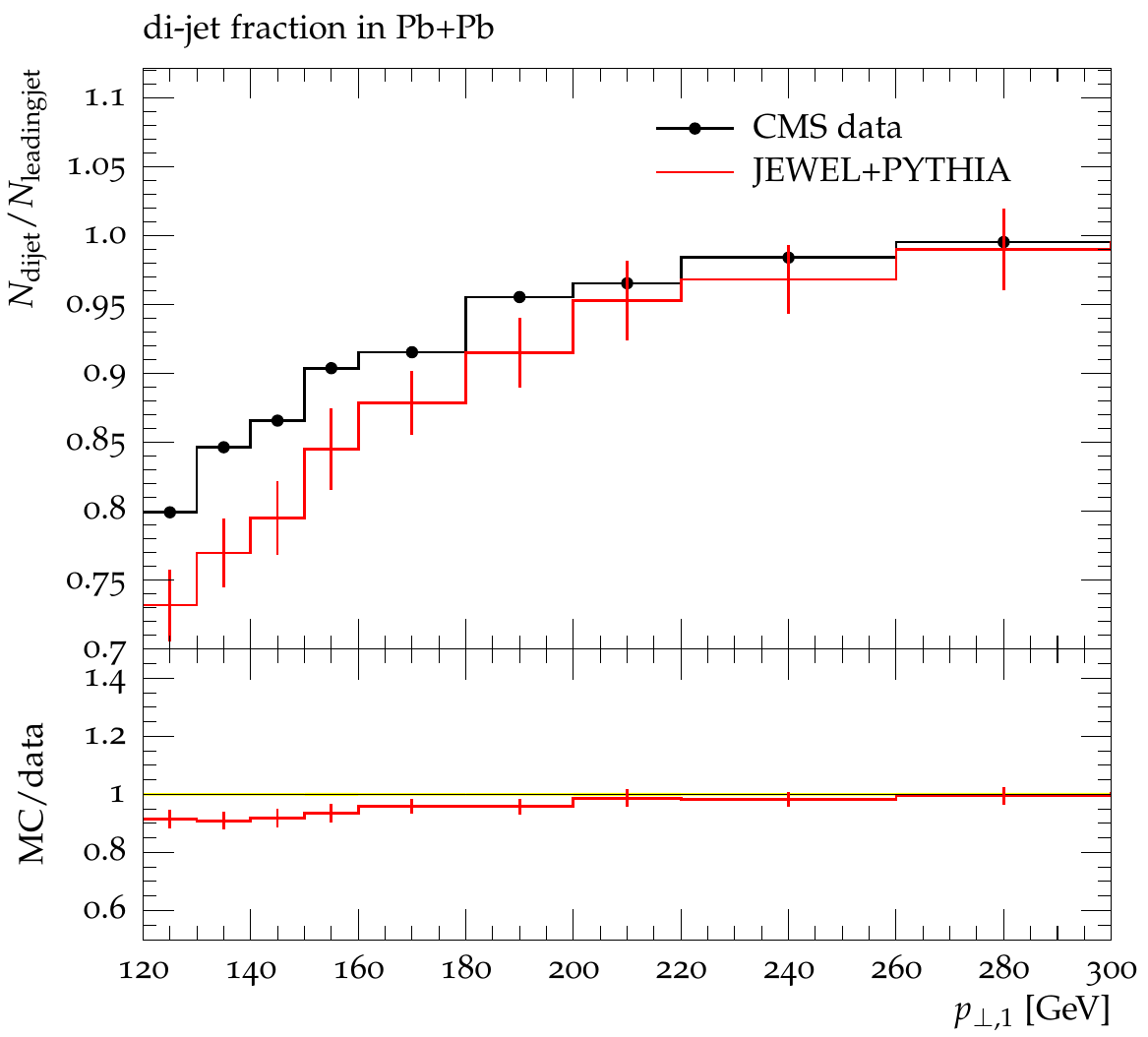}
\caption{Fraction of leading jets with a sub-leading jet fulfilling $p_{\perp,2} > \unit[30]{GeV}$ and $\Delta \phi > 2\pi/3$ as a function of the leading jet’s transverse momentum in p+p and Pb+Pb (\unit[0-20]{\%} centrality) collisions. \textsc{Jewel+Pythia} results are compared to CMS data~\cite{Chatrchyan:2012nia}. The data are not unfolded for jet energy resolution, so the Monte Carlo events were smeared with the parametrisation from~\cite{Chatrchyan:2012gt}.}
\label{fig::dijetfrac}
\end{figure*}

In the following measurements the azimuthal angle between the jets is required to be $\Delta \phi > 2\pi/3$, the problematic small $\Delta \phi > 2\pi/3$ region is thus excluded. The fraction of leading jets accompanied by a sub-leading jet fulfilling this requirement is shown in figure~\ref{fig::dijetfrac} as a function of the leading jet's transverse momentum. The agreement between data and \textsc{Jewel+Pythia} is excellent for p+p and slightly worse but still satisfactory in central Pb+Pb events.

\begin{figure*}[ht]
\centering
\includegraphics[width=0.45\linewidth]{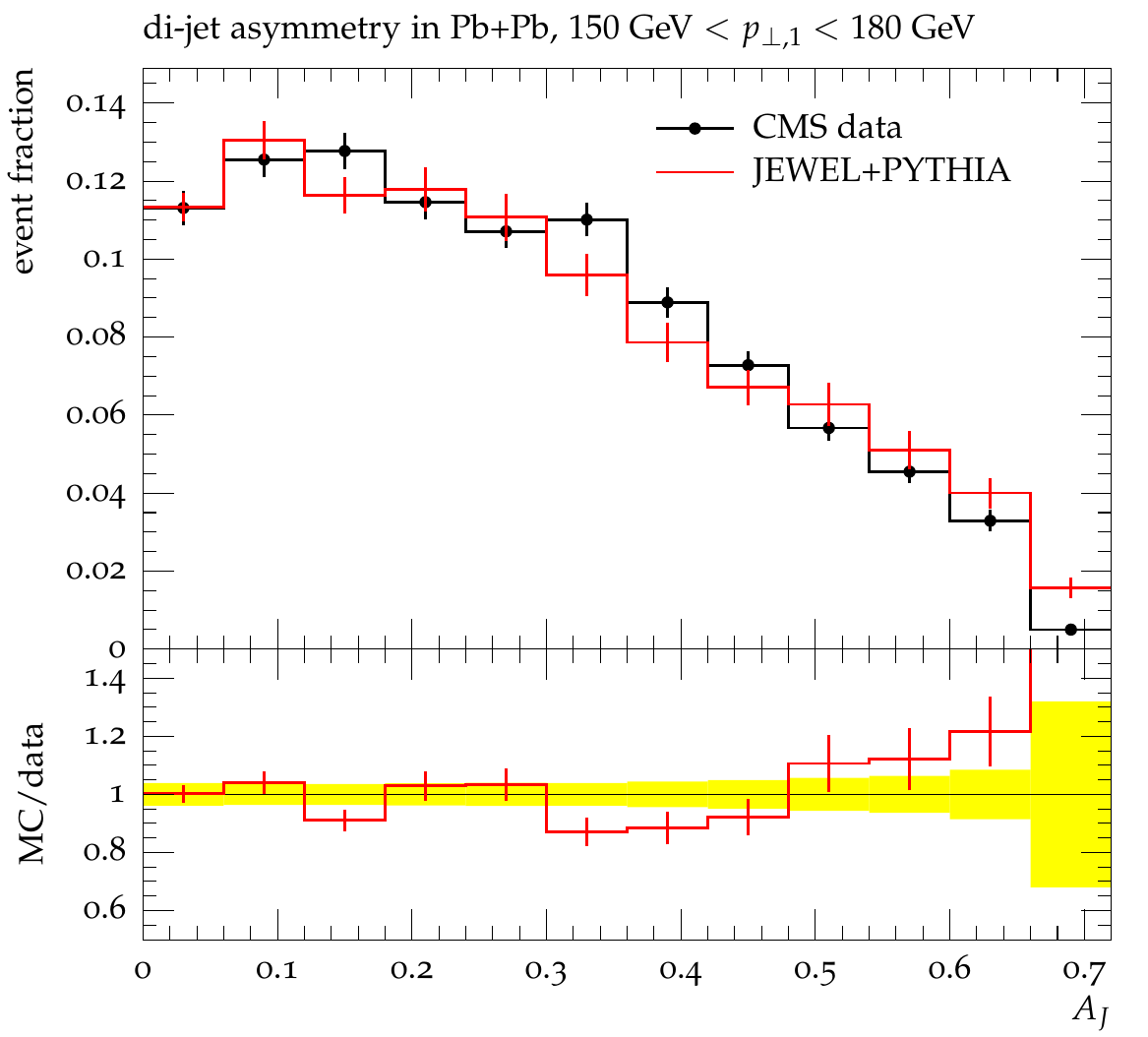}
\includegraphics[width=0.45\linewidth]{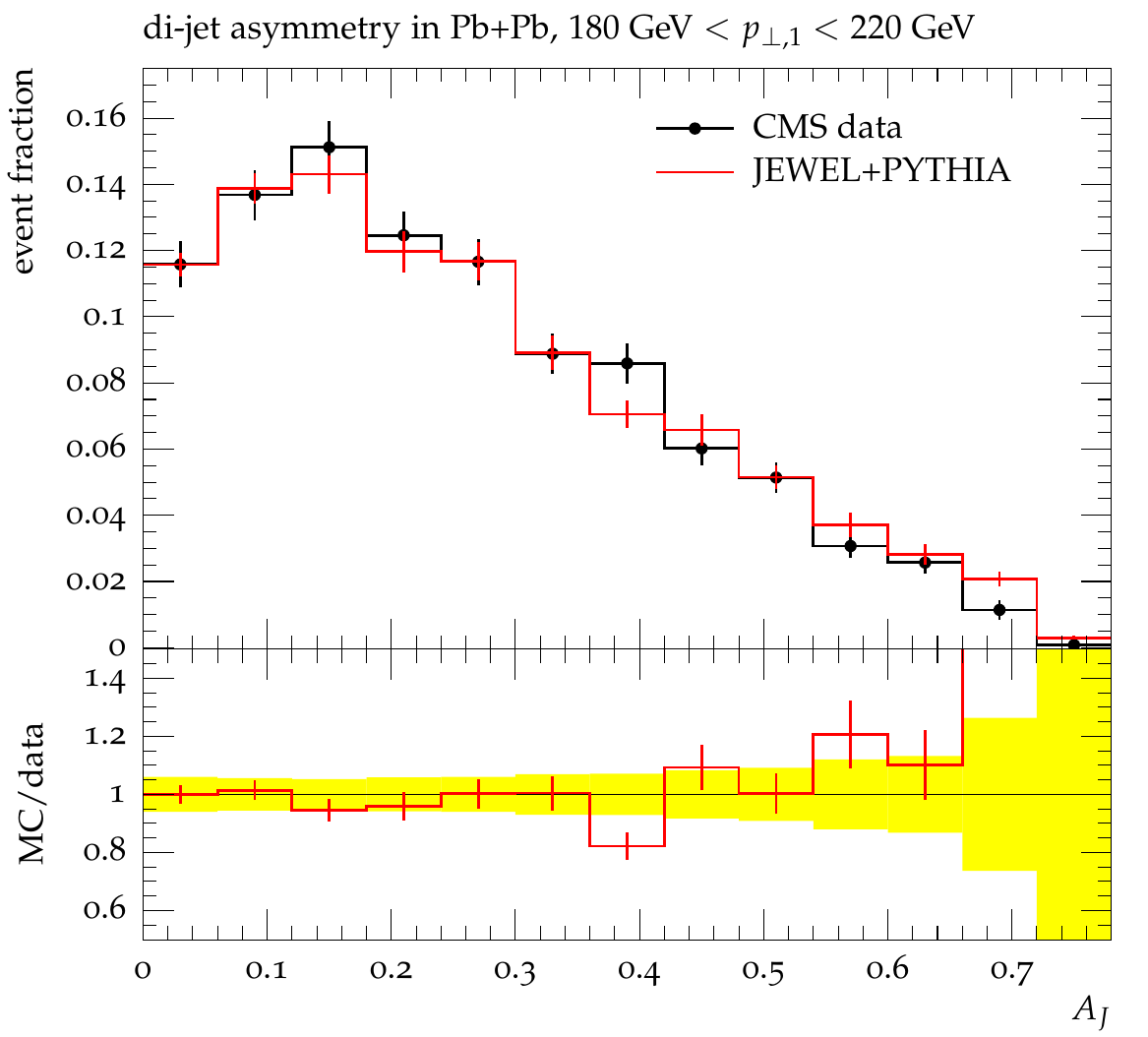}\par
\includegraphics[width=0.45\linewidth]{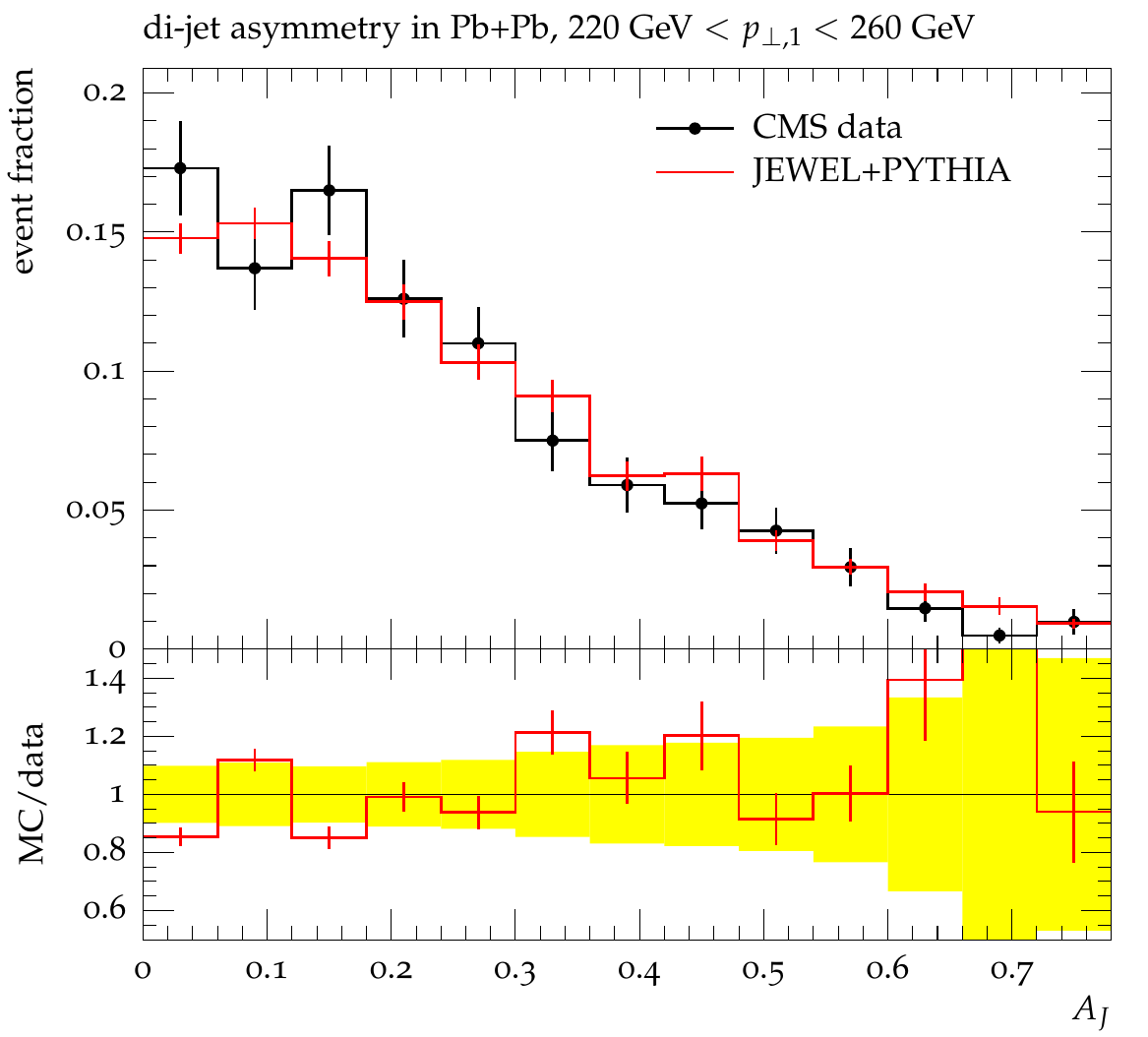}
\includegraphics[width=0.45\linewidth]{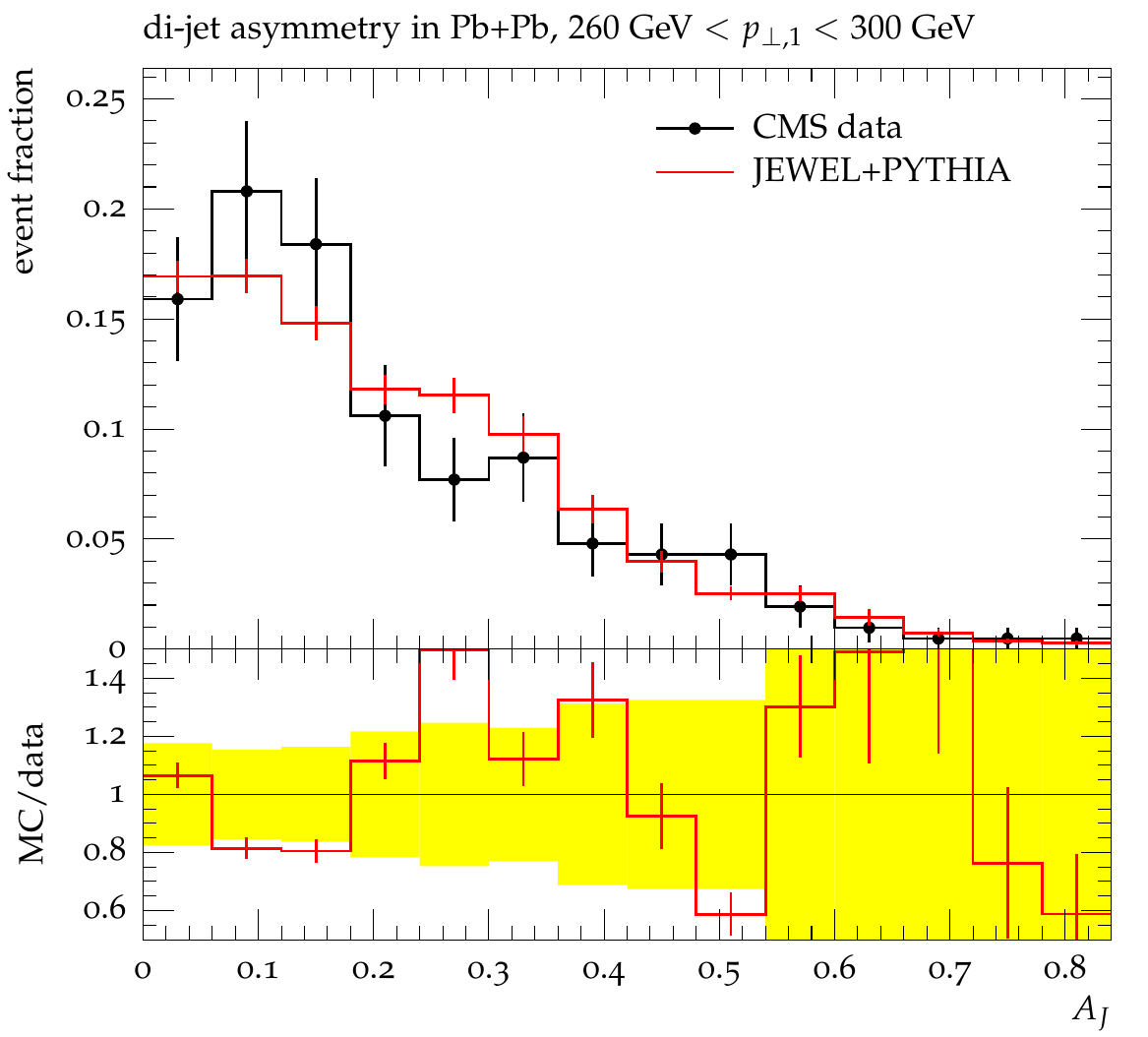}
\caption{Di-jet asymmetry $A_\text{J} = (p_{\perp,1} − p_{\perp,2}) / (p_{\perp,1} + p_{\perp,2})$ in Pb+Pb collisions (\unit[0-20]{\%} centrality) for different transverse momenta of the leading jet. The sub-leading jet is required to have $p_{\perp,2} > \unit[30]{GeV}$ and $\Delta \phi > 2\pi/3$. \textsc{Jewel+Pythia} results are compared to CMS data~\cite{Chatrchyan:2012nia}. The data are not unfolded for jet energy resolution, so the Monte Carlo events were smeared with the parametrisation from~\cite{Chatrchyan:2012gt}.}
\label{fig::aj}
\end{figure*}

\begin{figure*}[ht]
\centering
\includegraphics[width=0.45\linewidth]{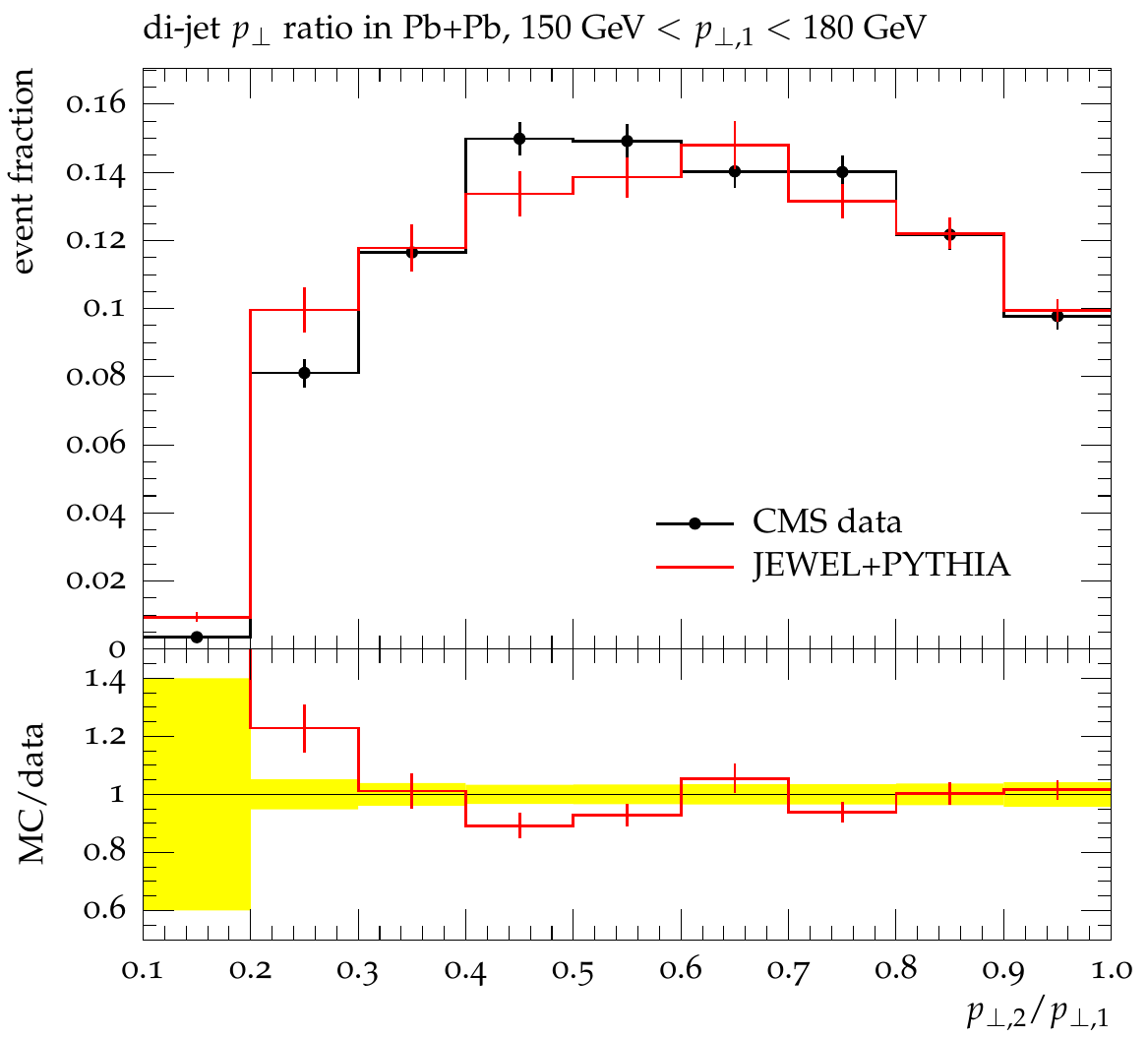}
\includegraphics[width=0.45\linewidth]{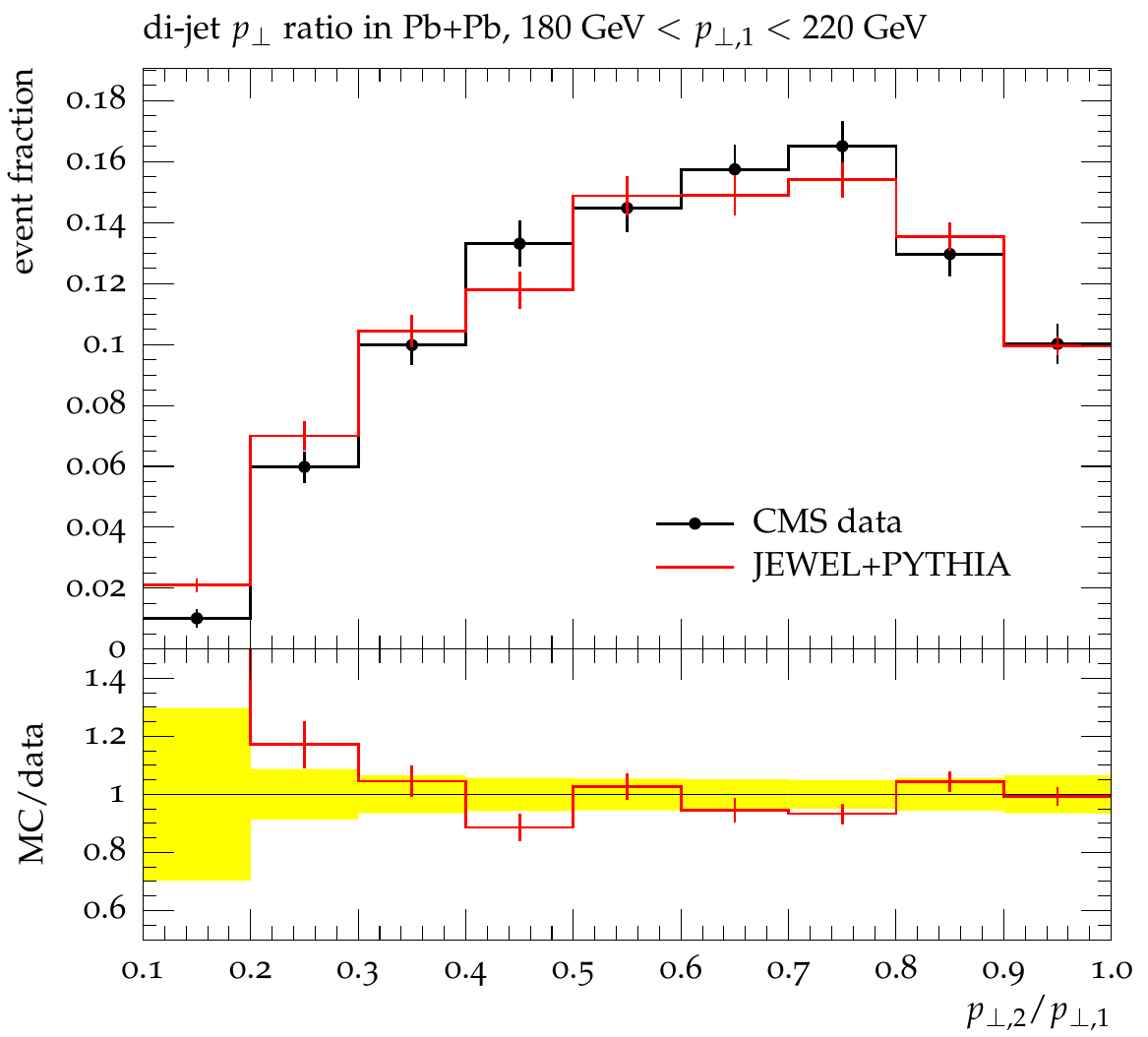}\par
\includegraphics[width=0.45\linewidth]{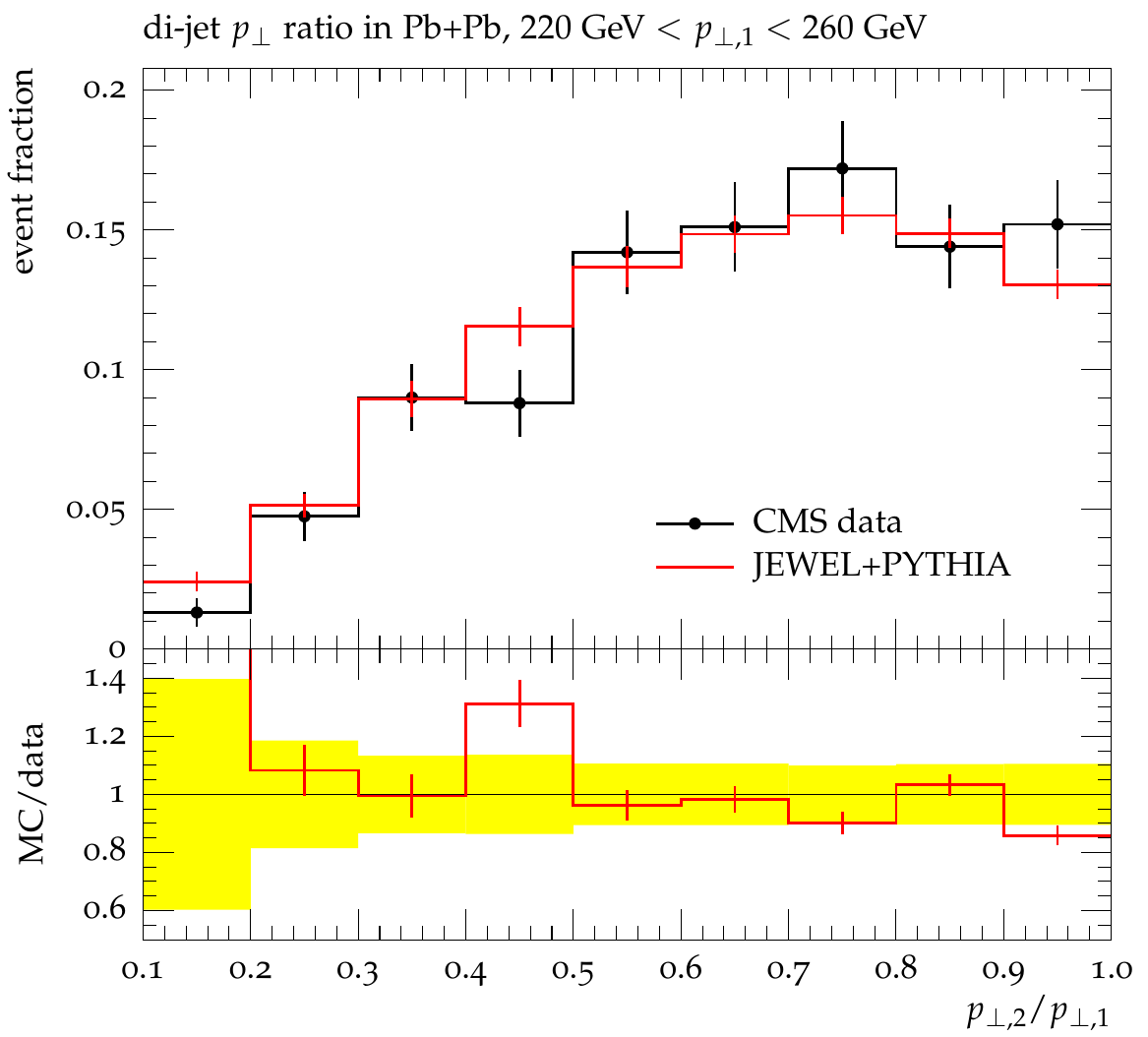}
\includegraphics[width=0.45\linewidth]{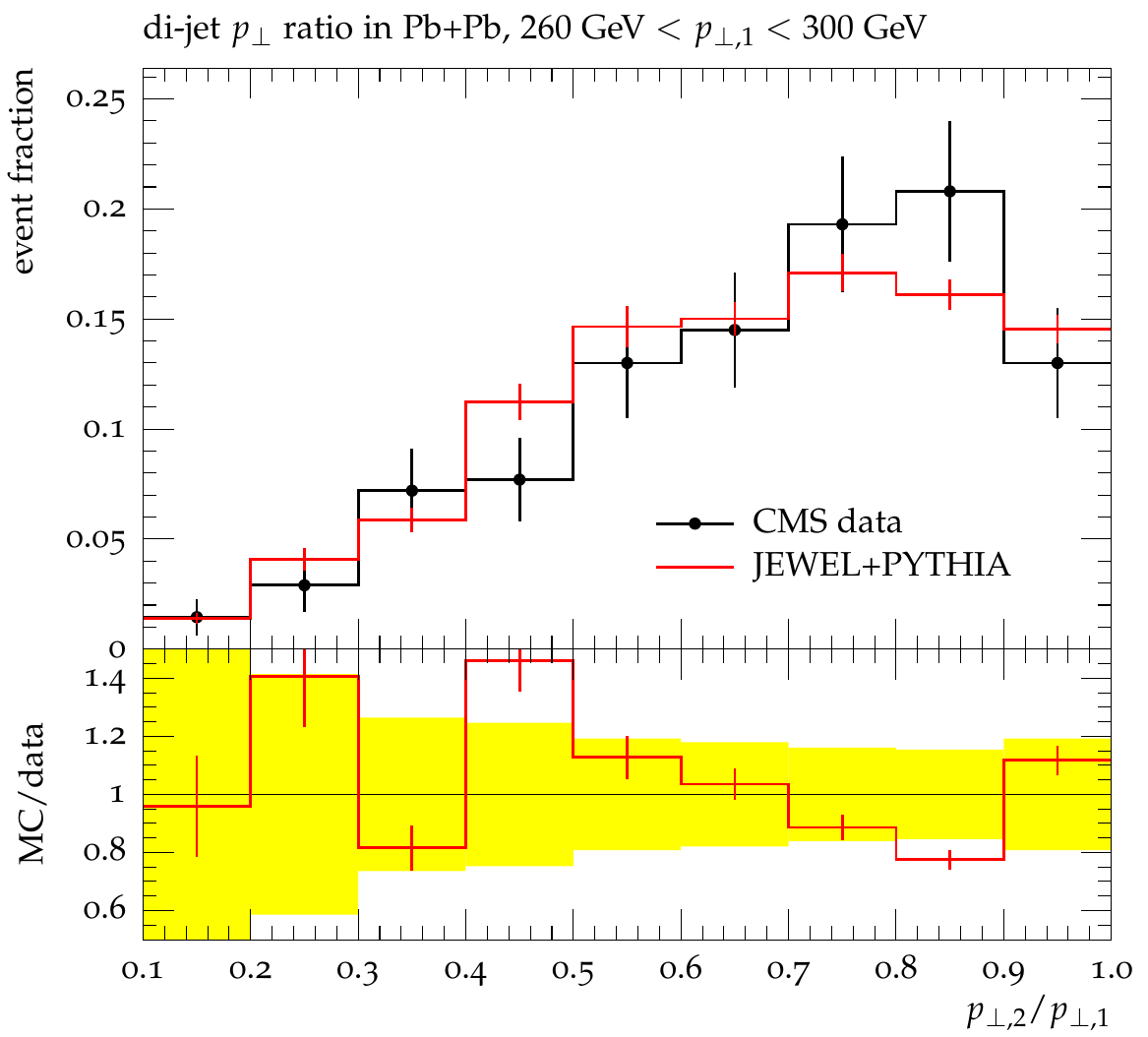}
\caption{$\pt$-ratio in di-jet events in Pb+Pb collisions (\unit[0-20]{\%} centrality) for different transverse momenta of the leading jet. The sub-leading jet is required to have $p_{\perp,2} > \unit[30]{GeV}$ and $\Delta \phi > 2\pi/3$. \textsc{Jewel+Pythia} results are compared to CMS data~\cite{Chatrchyan:2012nia}. The data are not unfolded for jet energy resolution, so the Monte Carlo events were smeared with the parametrisation from~\cite{Chatrchyan:2012gt}.}
\label{fig::ptratio}
\end{figure*}

\begin{figure*}[ht]
\centering
\includegraphics[width=0.45\linewidth]{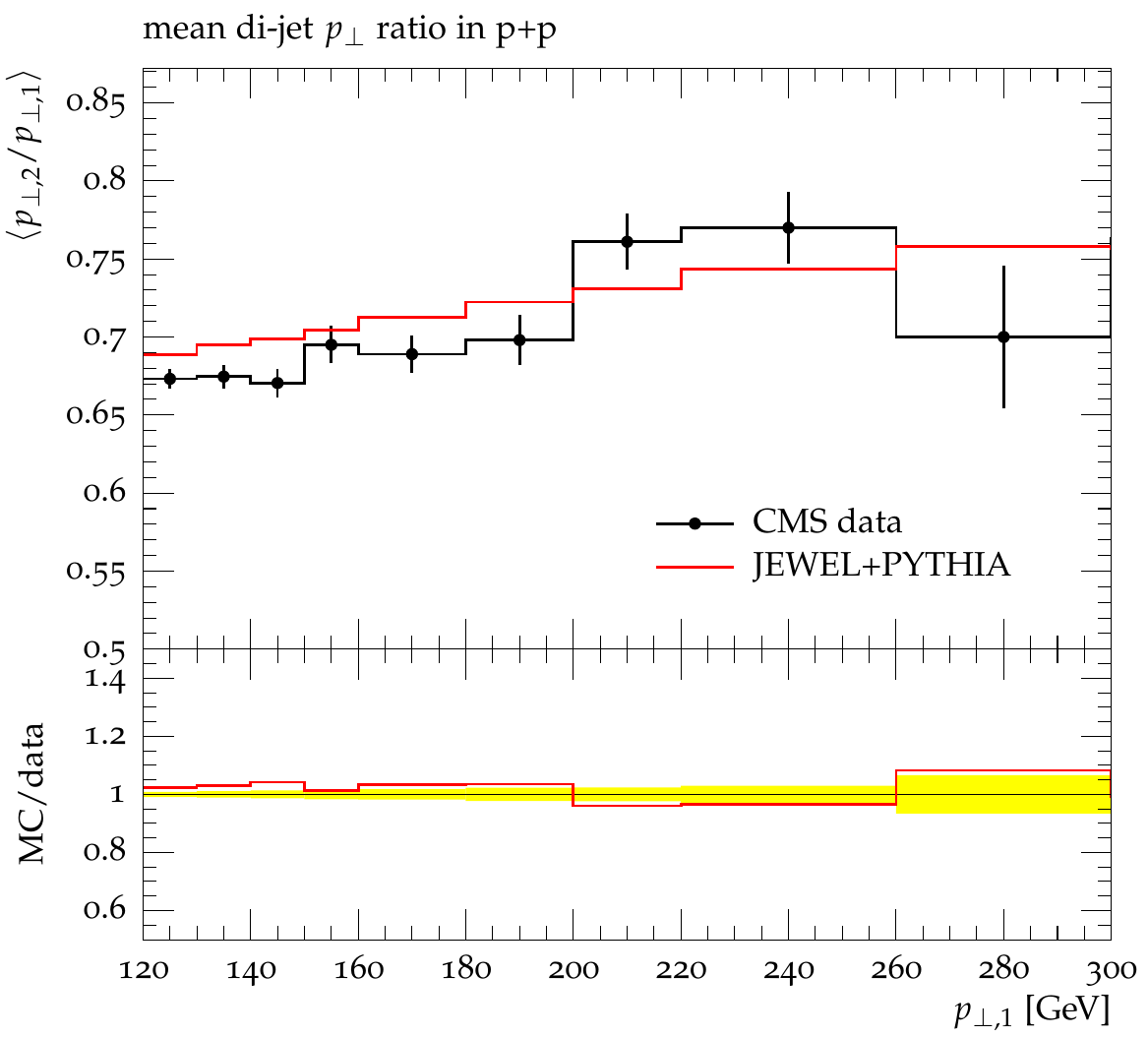}
\includegraphics[width=0.45\linewidth]{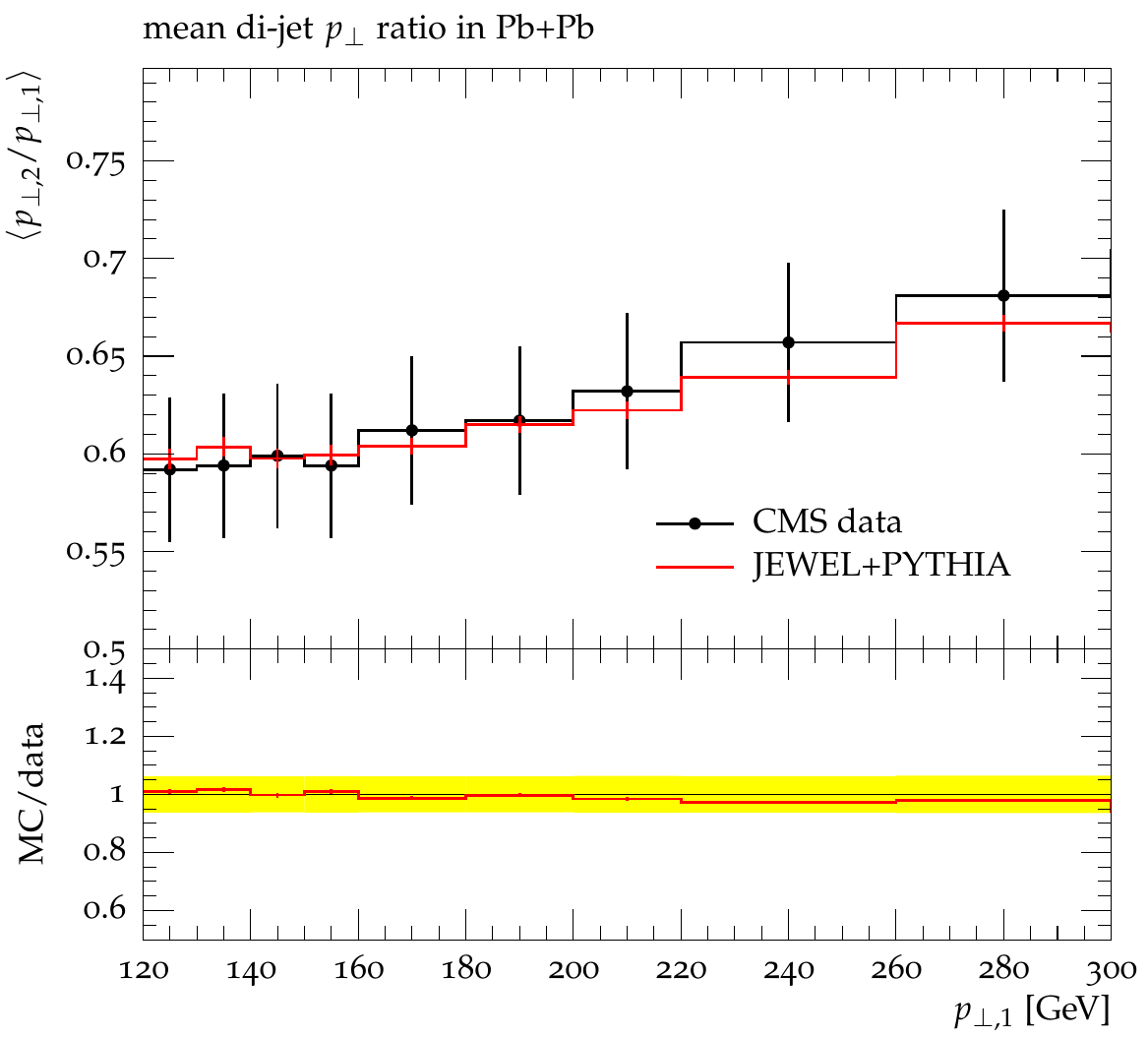}
\caption{Mean $\pt$-ratio in di-jet events in p+p and Pb+Pb collisions (\unit[0-20]{\%} centrality) as a function of the transverse momentum of the leading jet. The sub-leading jet is required to have $p_{\perp,2} > \unit[30]{GeV}$ and $\Delta \phi > 2\pi/3$. \textsc{Jewel+Pythia} results are compared to CMS data~\cite{Chatrchyan:2012nia}. The data are not unfolded for jet energy resolution, so the Monte Carlo events were smeared with the parametrisation from~\cite{Chatrchyan:2012gt}.}
\label{fig::meanptratio}
\end{figure*}

The asymmetry in di-jets is further quantified by the variable $A_\text{J} = (p_{\perp,1} - p_{\perp,2}) / (p_{\perp,1} + p_{\perp,2})$ (figure~\ref{fig::aj}) and the ratio $p_{\perp,2}/p_{\perp,1}$ of the transverse momenta of the jets (figure~\ref{fig::ptratio}). Both distributions are very well reproduced by \textsc{Jewel+Pythia}. It is thus not surprising, that the mean $\pt$-ratio shown in figure~\ref{fig::meanptratio} is also in excellent agreement. 

\begin{figure*}[ht]
\centering
\includegraphics[width=0.45\linewidth]{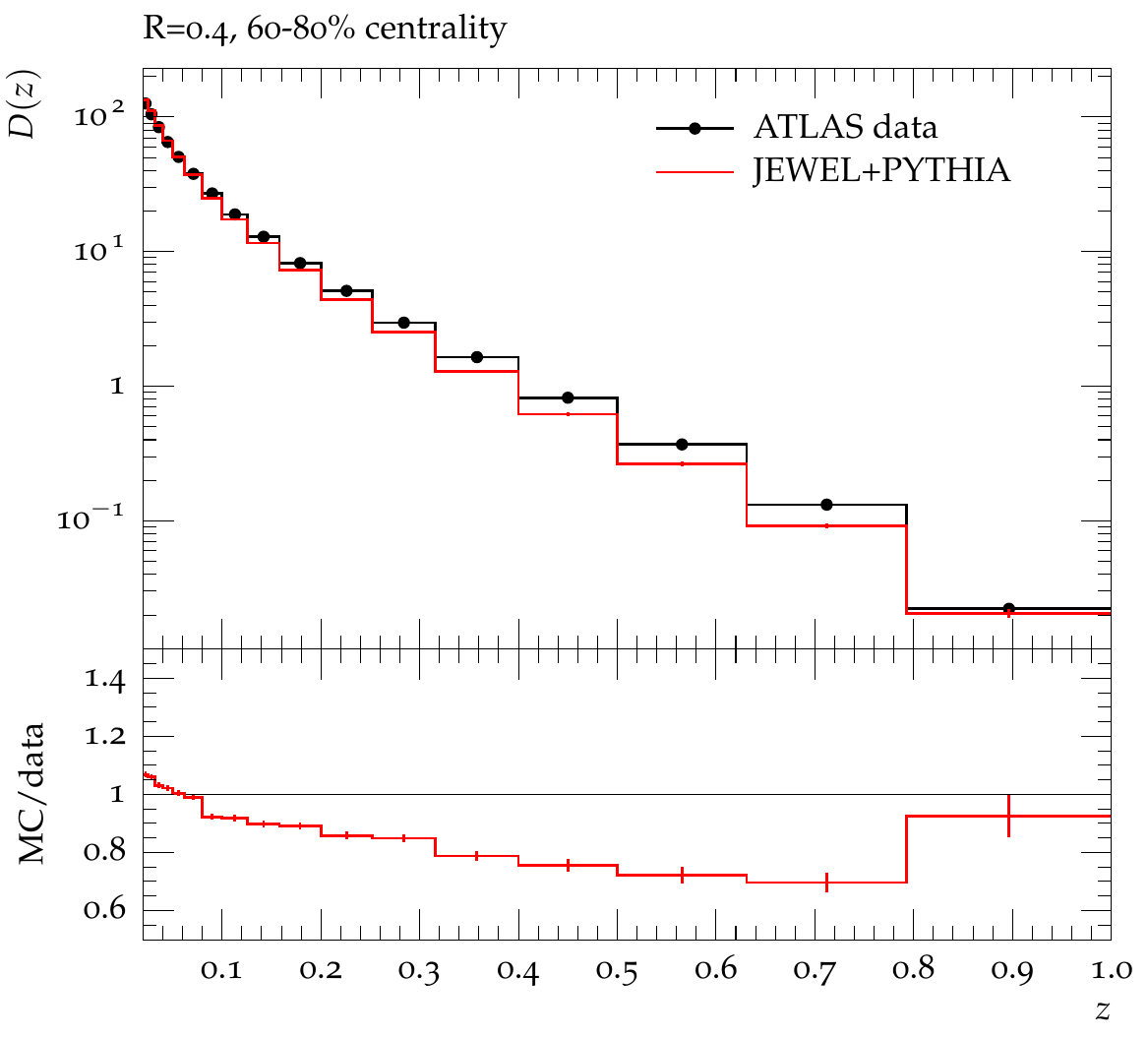}
\includegraphics[width=0.45\linewidth]{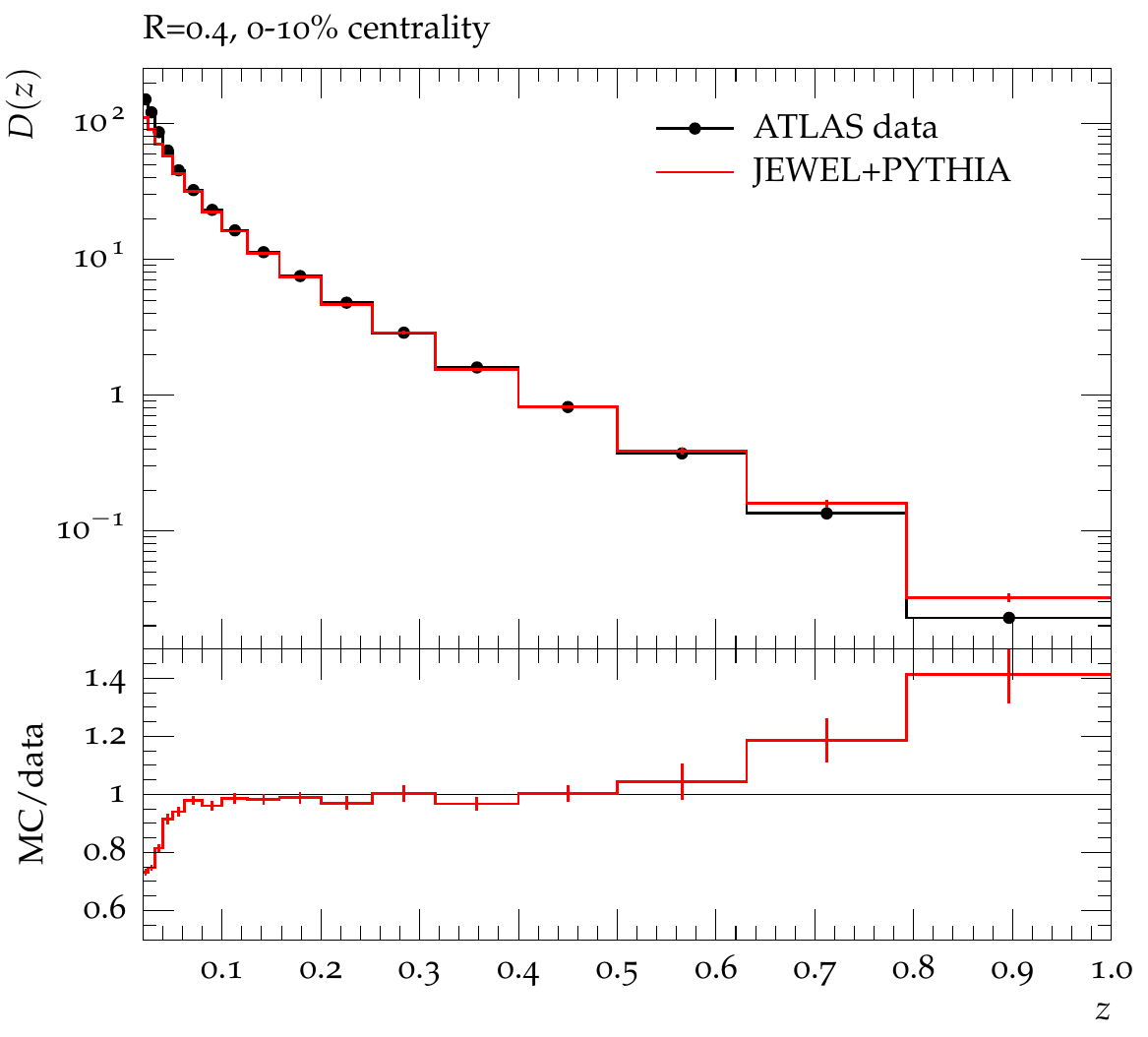}\par
\includegraphics[width=0.45\linewidth]{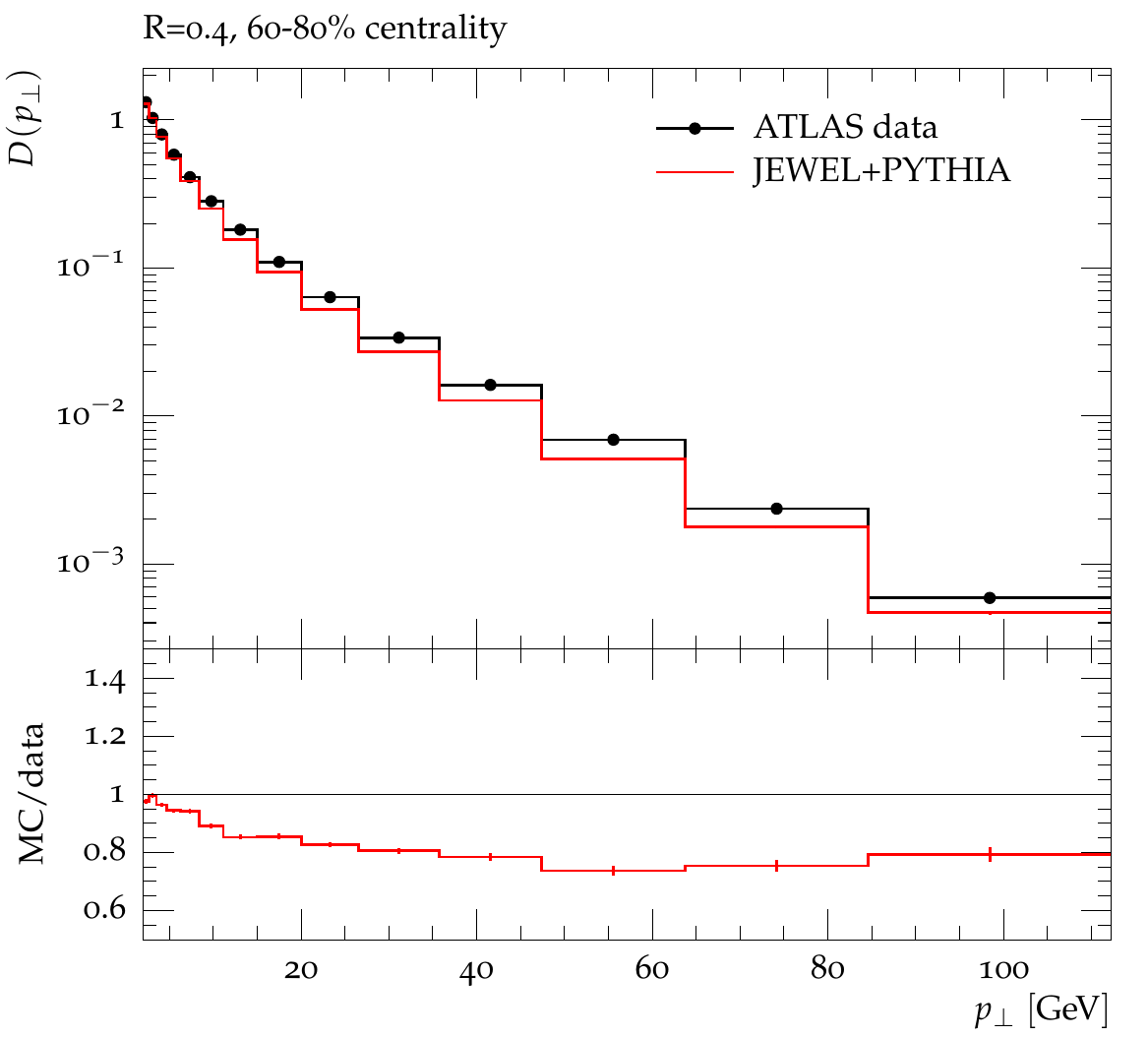}
\includegraphics[width=0.45\linewidth]{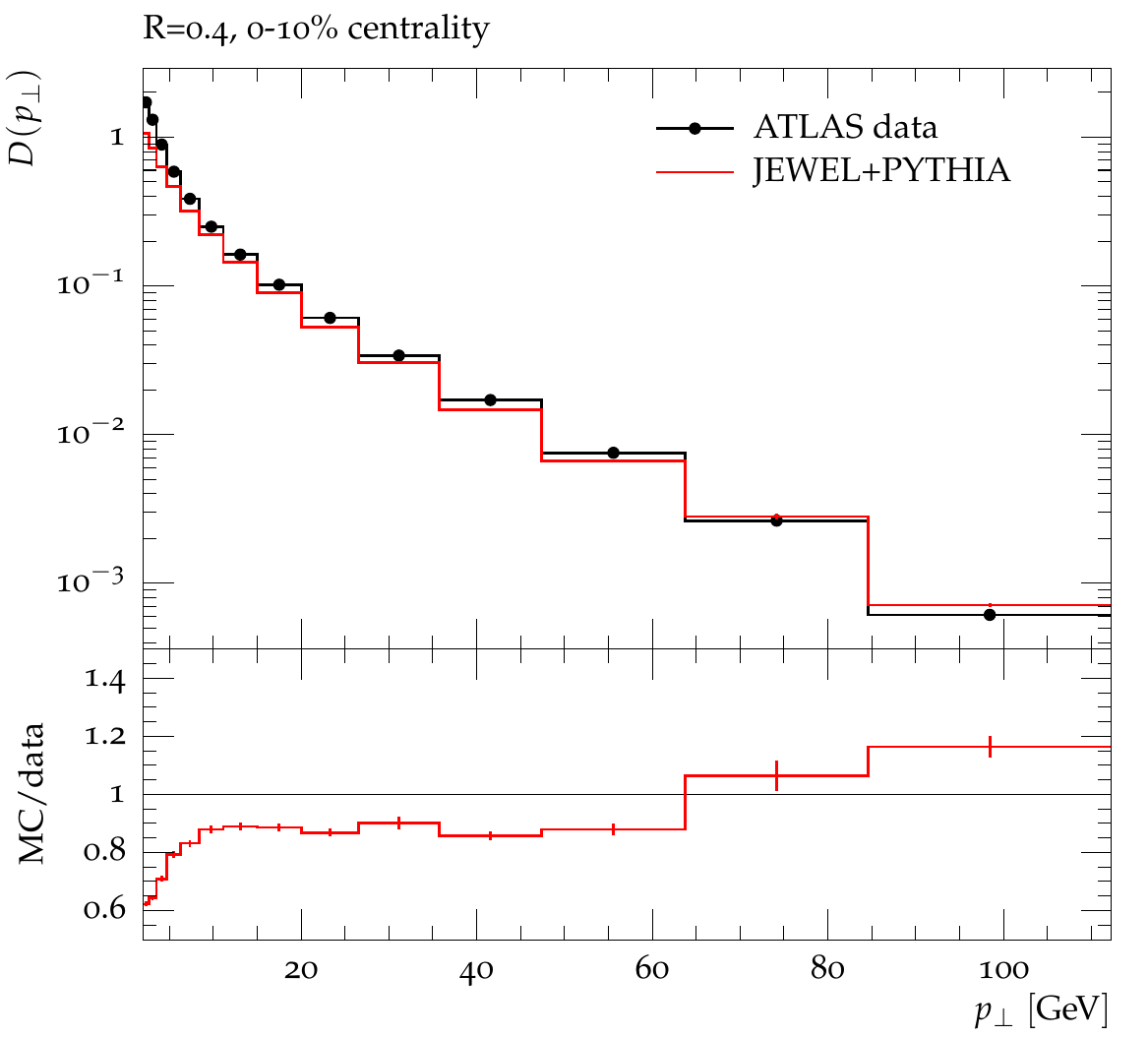}
\caption{\textsc{Jewel+Pythia} results for the fragmentation functions $D(z)$ (top) and $D(\pt)$ (bottom) for a jet radius of R = 0.4 in peripheral and central Pb+Pb events compared to ATLAS data~\cite{ATLAS-CONF-2012-115} (data points read off the plots, no errors shown).}
\label{fig::FFs}
\end{figure*}

\begin{figure*}[ht]
\centering
\includegraphics[width=0.45\linewidth]{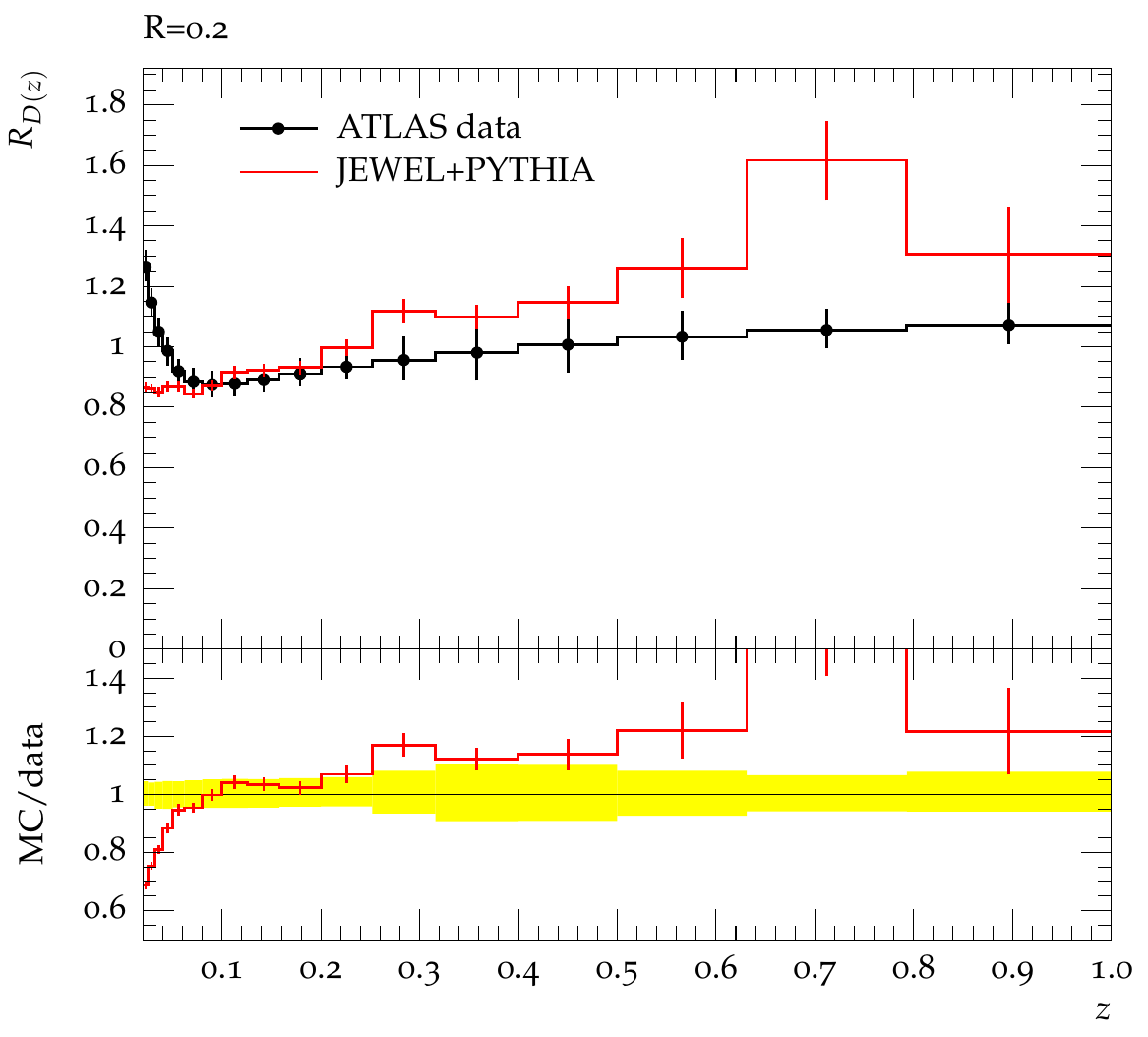}
\includegraphics[width=0.45\linewidth]{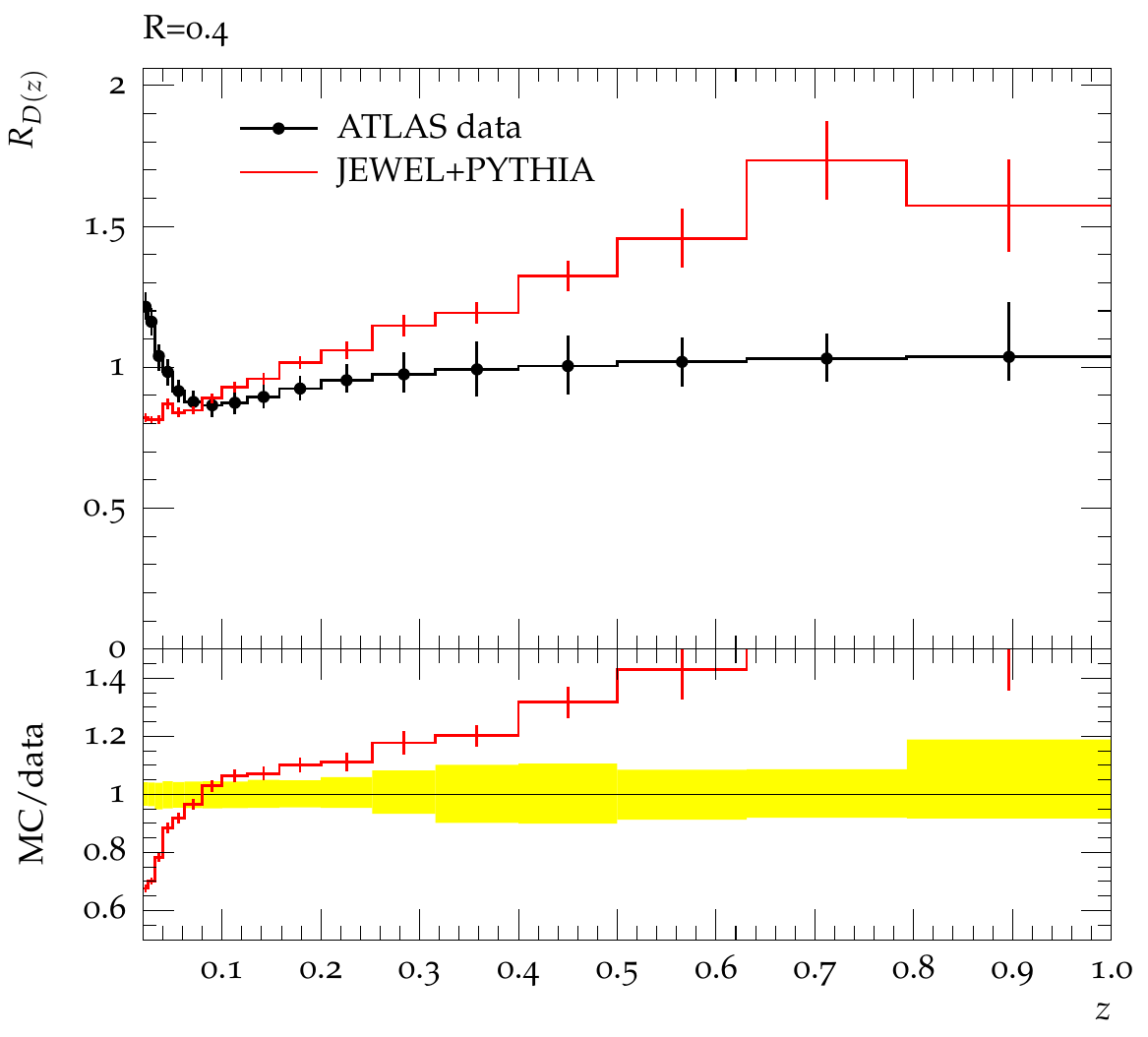}\par
\includegraphics[width=0.45\linewidth]{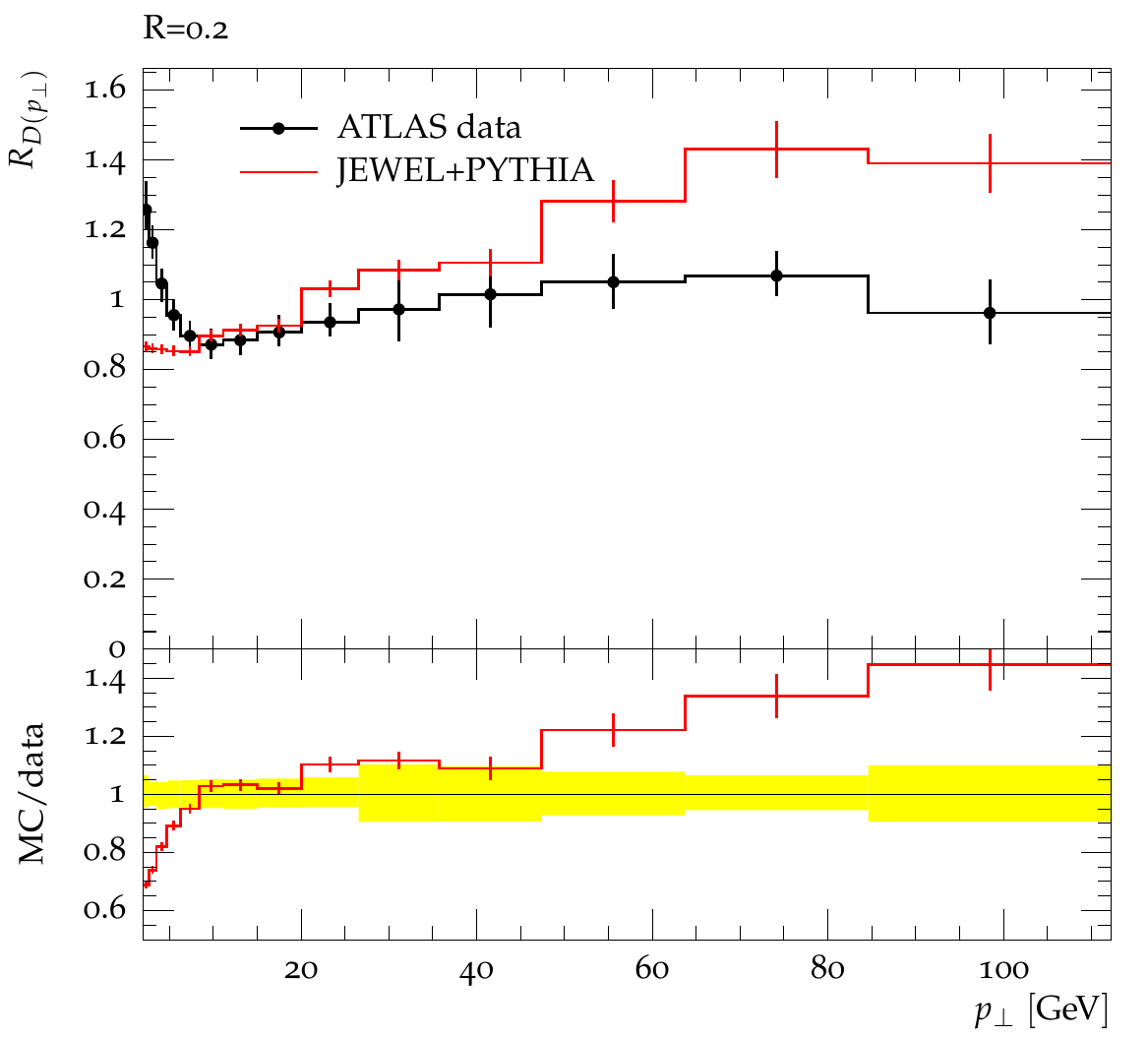}
\includegraphics[width=0.45\linewidth]{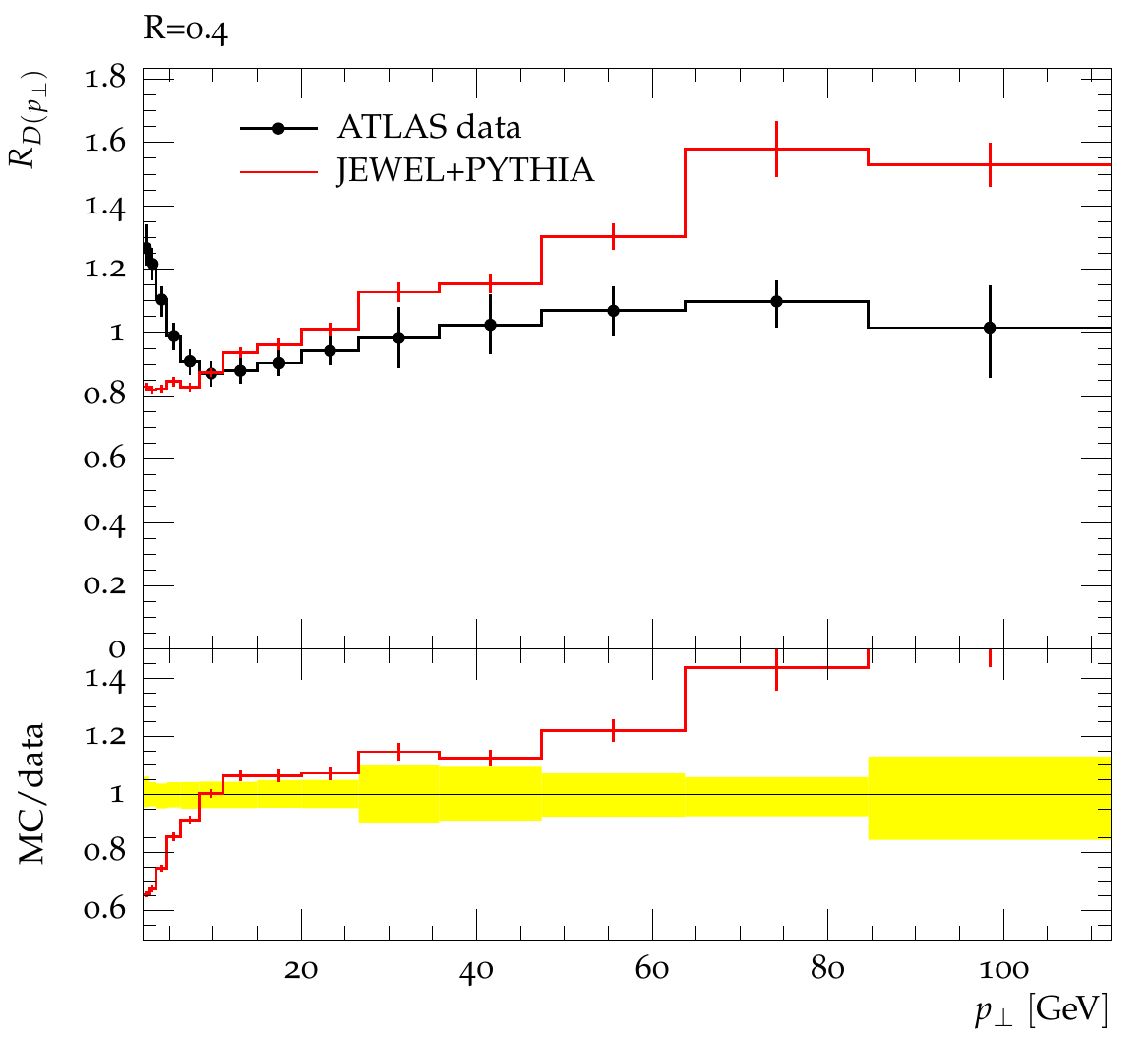}
\caption{\textsc{Jewel+Pythia} results for the ratios of the fragmentation functions $D(z)$ (top) and $D(\pt) $ (bottom) between central and peripheral Pb+Pb events for jet sizes $R=0.2$ (left) and $R=0.4$ (right) compared to ATLAS data~\cite{ATLAS-CONF-2012-115} (data points read off the plots, only maximum of statistical and systematic errors shown).}
\label{fig::FFratios}
\end{figure*}

Figure~\ref{fig::FFs} shows the intra-jet charged particle fragmentation functions as functions of the longitudinal momentum fraction $z$ and the transverse momentum in central and peripheral Pb+Pb collisions. The agreement between the \textsc{Jewel+Pythia} results and the ATLAS data is overall reasonable. The very low $z$/$\pt$ region is particularly sensitive to details of the modelling (e.g. the treatment of recoils) and \textsc{Jewel+Pythia} cannot be expected to describe it very well, in particular when the recoiling scattering centres are not kept in the event. There is a tendency in the Monte Carlo to fragment somewhat too soft in peripheral collisions, which is also observed in p+p events~\cite{Zapp:2012ak}. Consequently, the ratio of the fragmentation functions rises slighly while it stays flat in the data (figure~\ref{fig::FFratios}). This can happen since the hard core of the jets is protected from medium modifications due to the large scales involved in its formation. When the total momentum of the jet is reduced the fragmentation function becomes harder.

The overall agreement of \textsc{Jewel+Pythia} with the large variety of data is satisfactory, in particular since they were obtained with a rather simple model of the medium. A discussion of uncertainties can be found in~\cite{Zapp:2012ak}. 

\section{Running the code}

\subsection{Installation}

The medium is kept separate from the main program so that different models of the medium can be selected by linking to different medium codes. To simulate jet evolution in vacuum one also has to link to a medium model, but this one simply tells the \textsc{Jewel} that there is no medium.

\textsc{Jewel} relies heavily on \textsc{Pythia}\,6, for instance to simulate the matrix elements, initial state parton showers and hadronisation. It needs, however, a slightly modified version of \textsc{Pythia}\,6.4.25, which is distributed with the \textsc{Jewel} code and is not an official \textsc{Pythia} release. The modifications to the original \textsc{Pythia} code are (i) an enlarged event record (it has 23000 lines instead of 4000 in the modified version) to accomodate the larger heavy ion events, (ii) a slight extension of the LHAPDF interface which allows to use the EPS09 nuclear PDF sets and (iii) a customised PYEVWT routine that multiplies the differential cross section with a power of the parton $\pt$ to allow for the generation of weighted events.

The PDFs are loaded via \textsc{Pythia}'s LHAPDF interface and therefore a LHAPDF~\cite{Whalley:2005nh} installation is required. \textsc{Jewel} supports the use of the EPS09LO~\cite{Eskola:2009uj} nuclear PDF sets. The path to  LHAPDF has to be set in the Makefile for \textsc{Jewel}. For the default set-up of \textsc{Jewel} the CTEQ6LL and EPS09LOR\_208 PDF sets are required.

Compiling and the linking \textsc{Jewel} using the provided makefile results in two executables, namely \texttt{jewel-2.0.0-vac} and \texttt{jewel-2.0.0-simple}. The former simulates jet evolution in vacuum, the latter in a simple medium (cf.\ section~\ref{sec::medmodel}). Both have the name of a parameter file (cf.\ section~\ref{sec::jewelparams}) as optional argument. 

\subsection{Structure of the event generation}

\textsc{Jewel} does not simulate complete heavy ion events, but only the evolution of a di-jet system. First, \textsc{Jewel} initialises the geometric aspects, i.e. impact parameter and jet production point, of the event. Then the jet production matrix elements and initial state shower are generated by \textsc{Pythia}\,6.4~\cite{Sjostrand:2006za}. The proton PDFs are loaded via the LHAPDF interface, for the simulation of jet evolution in heavy ion collisions the EPS09 nuclear PDF set can be used on top of the selected proton PDF. The final state parton shower including possible interactions in a medium is generated by \textsc{Jewel}. The colour strings are also constructed by \textsc{Jewel} prior to hadronisation. There are two options how the colour can be arranged. One is to keep the colour topology essentially as in vacuum and treat recoils as if they were emissions~\cite{Zapp:2012ak} and the other model builds strings based on a criterion of minimal invariant mass~\cite{Zapp:2008gi}. After the strings have been constructed the event is handed back to \textsc{Pythia} for hadronisation and hadron decays. The conversion into HepMC\,2 events, finally, happens again in \textsc{Jewel}.

\subsection{The event format}

\textsc{Jewel} uses \textsc{Pythia}'s event record, which has been enlarged to 23000 lines. As heavy ion events can get very busy and to keep the events small, all intermediate particles are cleared from the event record before hadronisation. The events are written out in HepMC\,2 ascii format~\cite{Dobbs:2001ck}. Only the hadronic stage is written out, i.e. in $pp$ events the first vertex has the two beams as incoming particles and all primary hadrons (hadrons from string decays) as outgoing particles. In $e^+e^-$ in addition the decay of the virtual photon into quark-antiquark pair is written out explicitely to allow flavour specific analyses, the quark pair then decays into the primary hadrons. In both cases all subsequent hadron decays are contained in the event. To save disc space one can also choose to write out only the stable final state particles. For un-hadronised (partonic) events only this compressed output is currently available.

\subsection{Integration results}

During event generation integrals of the splitting functions, partonic PDFs 
and scattering cross sections are needed. As the numerical integration is 
costly in terms of computing time they are integrated at the beginning of the 
run and stored in tables. To save time, these tables are stored in files and 
can be read in from there in later runs. The integration results depend on the strong coupling $\alphas$, the parton shower cut-off $Q_0$, the medium parameters, $\sqrt{s}$ and the $\pt$ range in which jets are generated. The filenames for the three types are parameters of the main program. If a file of the given name exists, the results will be read in from there. If the files don't exist the code will do the integration and create the files to store the results. The program performs no checks to make sure that the integration results make sense for the chosen parameters of the run. It is thus the users responsability to ensure that the integration results and the parameter set are compatible.

\subsection{Treatment of recoiling scattering centres}

Normally, \textsc{Jewel} keeps the recoiling scattering centres in the event. This is the most natural thing to do for observables like single-inclusive hadron spectra. For observables that involve subtracting background, there is a problem. Since \textsc{Jewel} does not simulate the entire event, it is impossible to follow exactly the experimental procedure when analysing MC events. \textsc{Jewel} has the option to remove the recoiling scattering centres from the event before hadronisation. This leaves ambiguities when comparing to data, especially at low $\pt$ and for mixed observables that perform a background subtraction only for a part of the event. A satisfactory solution would require simulating the entire event including the reaction of the medium to the passage of a jet, which is currently not within reach.

\smallskip

Keeping the recoils in the event drastically increases the multiplicity and can lead to overflow of the event record. When this happens the event is discarded. \textsc{Jewel} has the option to suppress information about intermediate states in the event record to facilitate the handling of higher mulitplicity events. It is recommended that this option is enabled when the recoils are kept. High multiplicity events can also burst the bonds of the event record during the hadronisation and hadron decay stages simulated by \textsc{Pythia}, in this case \textsc{Pythia} will discard the current event. Obviously, when too many events are discarded this introduces a bias in the surviving sample. In the nucleus-nucleus centre-of-momentum frame, in which the simulation is performed, the density of scattering centres increases strongly with rapidity\footnote{In the co-moving frame the density decreses with increasing rapidity, but in the lab frame in increases due to the Lorentz contraction of the volume element.}. It is therefore advisable to restrict the rapidity region in which a medium is simulated as far as possible (cf.~section~\ref{sec::pscuts}).

\subsection{Phase space restrictions}
\label{sec::pscuts}

The phase space in which the jets are generated is restricted by imposing a minimal and a maximal $\pt$ on the hard matrix element. The lower cut is imperative due to the infra-red divergence of the matrix element and \textsc{Jewel} makes sure that $p_{\perp,\,\text{min}} \ge 3$\,GeV overwriting the user-defined parameter when necessary. The rapidity of the produced jets is unrestricted, but the medium is only simulated in a window around mid-rapidity (outside this window the density vanishes). Users should always be careful to generate events in a phase space that is sufficiently larger than the phase space in which their analyses operate. In particular, jets can not only migrate from larger to smaller $\pt$ due to medium interactions and incomplete reconstruction of the jet energy/momentum when using small and intermediate jet radius parameters, but there is also a finite probability for jets to gain energy in interactions with the medium. This effect is sizeable only for very small jet energies of the order of a few GeV, but can lead to large fluctuations because, although the probability for gaining energy may be small, low $\pt$ jets are produced with large weights or, when generating unweighted events, with high probability. The effect of the lower $\pt$ cut-off will thus be visible even above the cut-off. 

The rapidity region of the medium should also be chosen larger than the analysis region, because interactions in the medium allow jets and partons inside jets to migrate in rapidity and recoils can show up at relatively large distance from the jet.

\subsection{Parameters of JEWEL}
\label{sec::jewelparams}

When \textsc{Jewel} is executed from the command line the name of a parameter file can be passed as an optional argument. If no filename is provided \textsc{Jewel} will run with the default setting for \textsc{Jewel} and the medium model. The default setup is the one with which the results shown in section~\ref{sec::results} were obtained (except for the centrality, for which different choices were needed).
In the parameter file only the parameters with values deviating from the defaults have to be specified. Lines starting with a hash are interpreted as comments and skipped when reading in the parameters. The format of the other lines is <parameter name>\textvisiblespace<value> with only one parameter per line. A complete list of the \textsc{Jewel} parameters with their default values in parantheses is given below. The medium model and its parameters are kept separat from the main program and are explained in section~\ref{sec::medmodel}.

\begin{description}
\item[NEVENT\,(10000):] number of events to be generated
\item[NJOB\,(0):] arbitrary job number used to initialise the random number generator
\item[LOGFILE\,(`out.log'):] name of the log file
\item[HEPMCFILE\,(`out.hepmc'):] name of file to which events are written\footnote{This can also be the name of a fifo which can be used pass the events directly on to the analysis code.}
\item[SPLITINTFILE\,(`splitint.dat'):] name of file containing integrated splitting functions
\item[PDFFILE\,(`pdfs.dat'):] name of file containing integrated partonic PDFs
\item[XSECFILE\,(`xsecs.dat'):] name of file containing integrated scattering cross sections
\item[MEDIUMPARAMS\,(`medium-params.dat'):] config file for medium model
\item[NF\,(3):] number of flavours used to evaluate $\alphas$
\item[LAMBDAQCD\,(0.4):] $\Lambda_\text{QCD}$ [GeV]
\item[Q0\,(1.5):] infra-red parton shower cut-off [GeV]
\item[PTMIN\,(5.):] minimum $\pt$ in matrix element [GeV]
\item[PTMAX\,(350.):] maximum $\pt$ in matrix element [GeV] (inactive when PTMAX < 0)
\item[ETAMAX\,(3.1):] rapidity range [-ETAMAX, ETAMAX] in which a medium is simulated
\item[PROCESS\,(`PPJJ'):] process that is to be simulated by matrix element, currently available are di-jet production in $e^+e^-$ (`EEJJ') and $pp$ (`PPJJ') collisions
\item[SQRTS\,(2760.):] c.m.s. energy of the colliding system [GeV]
\item[PDFSET\,(10042):] LHAPDF number for the (proton) PDF set\footnote{see  \url{http://lhapdf.hepforge.org/pdfsets} for a list of available PDF sets and their numbers}
\item[NSET\,(1):] number of EPS09 nuclear PDF set (0: none, 1: central value, 2-31: error sets)
\item[MASS\,(208.):] mass number of nucleus (yes, it has to be a double)
\item[WEIGHTED\,(T):] switch for weighted/unweighted events
\item[WEXPO\,(5.):] for weighted events: power of $1/\pt$ with which to oversample
\item[ANGORD\,(T):] switch for angular ordering
\item[KEEPRECOILS\,(F):] switch for keeping recoiling scattering centres 
\item[HADRO\,(T):] hadronisation switch
\item[HADROTYPE\,(0):] type of colour arrangement (0: vacuum like, 1: model based on minimising invariant mass of strings)
\item[SHORTHEPMC\,(T):] compact event output containing only stable final state particles
\item[COMPRESS\,(T):] delete information about intermediate states from event record to allow generation of higher multiplicity events
\end{description}

\subsection{The medium model and its parameters}
\label{sec::medmodel}

The medium model is not part of the main \textsc{Jewel} code but has to be linked from a separate file. \textsc{Jewel} is shipped with a set-up for baseline calculations in vacuum, which obviously has no medium related parameters, and a simple medium model~\cite{Zapp:2005kt,Zapp:2012ak}. The latter is a Bjorken~\cite{Bjorken:1982qr} model describing the boost-invariant longitudinal expansion of an ideal quark-gluon gas. The density profile and other geometrical aspects such as the distribution of jet production points  are taken from a Glauber model~\cite{Eskola:1988yh}. The parameters of the medium model are read from a separate parameter file. If no parameter file is found the code will run with the default settings. Again, only parameters with values differing from the default have to be specified.

\begin{description}
\item[TAUI\,(0.6):] initial time $\tau_\text{i}$ [fm]
\item[TI\,(0.36):] (mean) initial temperature $T_\text{i}$ [GeV]
\item[TC\,(0.17):] critical temperature $T_\text{c}$ [GeV]
\item[WOODSSAXON\,(T):] switch between Woods-Saxon potential and hard sphere
\item[CENTRMIN\,(0.):] lower end of centrality range to be simulated [\%]
\item[CENTRMAX\,(10.):] upper end of centrality range to be simulated [\%]
\item[NF\,(3):] number of quark flavours in the quark-gluon gas
\item[A\,(208):] mass number of colliding nuclei (this needs to be an integer)
\item[N0\,(0.17):] density parameter of Woods-Saxon potential [fm$^{-3}$]
\item[D\,(0.54):] thickness parameter of Woods-Saxon potential [fm]
\item[SIGMANN\,(6.2):] nucleon-nucleon cross section [fm$^2$]
\item[MDFACTOR\,(0.45):] minimum of infra-red regulator [GeV] (has to be larger than $\Lambda_\text{QCD}$)
\item[MDSCALEFAC\,(0.9):] factor multiplying infra-red regulator, i.e. $\mu_\text{D}=3T$
\end{description}

\subsection{Interpreting the logfiles and error handling}
\label{sec::errors}

The logfile starts with the date and time at which the job started and the \textsc{Jewel} banner containing version, references etc. Then the parameters of the current runs are printed out followed by the \textsc{Pythia} banner. The code then reports whether the cross sections, PDFs and splitting functions are integrated or read from a file. This concludes the initialisation phase. 

During event generation \textsc{Jewel} prints a progress statement after every completed percent of the job (i.e.\ when generating 1000 events a message is printed every time 10 events have been completed). Both \textsc{Jewel} and \textsc{Pythia} also print warnings and error messages. The \textsc{Jewel} warnings mainly concern numerical precision and are harmless as long as they are not too frequent and/or the reported deviations are not large. In case of serious trouble \textsc{Jewel} discards the event and prints an error message. Again, this will happen from time to time (for instance when the event is too long) and is often harmless as long as it does not occur frequently. Failures inside \textsc{Pythia} are typically of a more serious nature. Therefore, when \textsc{Pythia} discards more than 5\,\% of the events the run is aborted. 

At the end of the run \textsc{Jewel} reports the mean number of splittings, (single) momentum transfers and effective momentum transfers. The ratio of the last two numbers thus indicates how many scattering centres on average act coherently. All these numbers are for illustrative purposes only and are not strictly physically meaningful. More important are the numbers of successful and discarded events. Obviously, when a large fraction of the events was discarded, one should not trust the remaining ones. In rare cases the conversion from the internal event format to hepmc fails, the number of these incidents is also given at the end of the logfile. Next, the generated di-jet cross section per proton-proton collision given the $\pt$ cuts is quoted together with the sum of all generated event weights. Since each event carries its own weight this number is only given for cross checks. Finally, the last line gives the date and time at which the job terminated.

\section{Acknowledgements}
Jochen Klein is thanked for helping in separating the medium model from the main program and the implementation of the simple medium model provided with \textsc{Jewel}. A routine for calculating the exponential integral by S.~Zhang and J.\,M.~Jin~\cite{SpecialFnc} is used by \textsc{Jewel}.

\section {Disclaimer}
The \textsc{Jewel} code is provided without any warranty under the terms of the GNU General Public License (GPL) version 2, users should be wary and use common sense when judging and interpreting their results. It is copyrighted but may be used for scientific work provided proper reference is given.

%

\bibliographystyle{k-physrev4}
\bibliography{jetquenching}

\end{document}